# UNDERSTANDING DOPING OF QUANTUM MATERIALS

Alex Zunger* and Oleksandr I. Malyi

Energy Institute, University of Colorado, Boulder, Colorado 80309

## Abstract

Doping mobile carriers into ordinary semiconductors such as Si, GaAs, and ZnO was the enabling step in the electronic and optoelectronic revolutions. The recent emergence of a class of "Quantum Materials", where uniquely quantum interactions between the components produce specific behaviors such as topological insulation, unusual magnetism, superconductivity, spin-orbit-induced and magnetically-induced spin splitting, polaron formation, and transparency of electrical conductors, pointed attention to a range of doping-related phenomena associated with chemical classes that differ from the traditional semiconductors. These include wide-gap oxides, compounds containing open-shell d electrons, and compounds made of heavy elements yet having significant band gaps. The atomistic electronic structure theory of doping that has been developed over the past two decades in the sub-field of semiconductor physics has recently been extended and applied to quantum materials. The present review focuses on explaining the main concepts needed for a basic understanding of the doping phenomenology and indeed peculiarities in quantum materials from the perspective of condensed matter theory, with the hope of forging bridges to the chemists that have enabled the synthesis of some of the most interesting compounds in this field.

## 1. INTRODUCTION

*Doping and doping bottlenecks:* Doping is the creation of ionizable entities within the solid, leading to mobile free carriers and the ensuing shift of the Fermi level $E_F$. Failure to position the Fermi level in a solid via doping at a target energy often disables carrier transport, superconductivity, topological insulation, and other effects predicated at the occupation of target energy levels. This article will explain how the "modern quantum theory of doping" developed in semiconductor physics demystified many peculiarities and misconceptions regarding doping bottlenecks. This will provide a better understanding of the main "design principles" needed for successful doping and point to strategies of carrier insertion in difficult to dope compounds, turning, for example, "Mott insulators" into "Mott semiconductors".

Doping can be attempted by *impurities*[1-3], or via the formation of carrier-producing *structural defects* (e.g., vacancies or anti-site defects)[4,5], or by direct carrier injection (gating)[6], or via the creation of a transient carrier concentration by photoexcitation ("photodoping")[7]. Alloying or plain chemical substitution that does not produce free carriers will not be considered here as doping. Impurities can either become an integral part of the skeletal host crystal structure (as in substitutional impurities in semiconductors[2]), or they can become more weakly interacting entities, not forming direct bonds with the host crystal as in interstitial alkali atom doping in organic compounds[8-10], or as in interstitial doping (e.g., hydrogen) in inorganic solids[11-16], or as in "modulation doping"[17], where the doped ions are confined to a neighboring layer.

Not all such chemical insertions result in doping because of a variety of "doping bottlenecks", e.g., (i) the impurity might have limited solid solubility, thus precipitate as a secondary phase without forming electrically active states[18]; what's worse, the selected dopant might be reactive, forming unwanted compounds with the host elements[19]; (ii) even if the impurity atom does enter the solid network, it might form an amphoteric center, creating both donor and acceptor levels as Cu in $Bi_2Se_3$[20] or Au in Ge[21] that compensate each other; (iii) even if the substituted impurity does form either a donor or an acceptor, its ionization energy level might be energetically too deep in the gap (with carriers strongly bound to the impurity as in Mott insulators[22]) preventing ionization at relevant temperatures; (iv) even if the impurity forms levels that are *energetically* not deep, their *wave functions* might be localized (e.g., forming localized polaronic states[23] that trap carriers), thus not contributing *free* carriers. Finally, (v) even if substitution does form initially mobile free carriers, the sheer existence of free carriers in a solid can induce a structural change in the solid (the "self-regulation response"[24-26]) that compensates the deliberately introduced free carriers, leaving the system carrier-free. Whereas obstacles (i)-(iv) can often be overcome by selecting different impurities or optimizing the growth method, the last mechanism (v) is often the final frontier, setting up insurmountable doping bottlenecks, because in this case, *the very existence of free carriers above a certain threshold causes their demise*. Thus, the wish often expressed in the literature to dope a material to a *given* Fermi level (by counting how many added carriers would be needed to place $E_F$ at a given energy) is often just wishful thinking, as one of the doping bottlenecks—often (v)—can set in, bringing doping to a halt. Indeed, the ability to create free carriers and an ensuing shift of the Fermi-level via doping is the defining aspect of a gaped solid being a semiconductor[27]; an insulator is then functionally a gapped solid that failed doping. Successful doping—such as the case in the classic semiconductors Si or GaAs is generally a rare event, in particular, for ambipolar doping of both for electrons (n-type) and holes (p-type).

*Relevance of understanding doping and doping bottlenecks*: Technologies that are based on carrier transport in gapped materials (e.g., transistors, light-emitting diodes, photovoltaic cells, and sensors) are rendered useless if the active material in them cannot be doped because of any of the mechanisms (i)-(v) above. Not just technology may be affected by failure to dope. There are cases where theoretically predicted interesting effects such as Rashba spin splitting[28], Dirac points in topological systems[29-32], band inversions in topological insulators[30-33], or quantum spin liquids that can model high-temperature superconductivity physics[34] could be explored only if the Fermi level coincides with a certain region in the band structure. Yet, to reach such points, one would often need to dope the material to reach that target Fermi level. This target may sometimes require significant shifts in $E_F$ that may not be granted by the host compound. Indeed, the material base amenable to transport technologies has been historically severely limited by the inability to induce free carriers in large classes of materials. Not surprisingly, a central question hovering over this field is *Are doping bottlenecks an intrinsic propensity of compounds encoded in their structure and electronic makeup or a temporary setback that can be overcome by clever growth tricks?* This is a consequential issue because the answer determines the essential strategy of the successful design of functional materials and the technologies based on them.

*New classes of Quantum Materials now call for a deeper understanding of doping*: Historically, the study of doping was limited to traditional semiconductors such as Si, Ge, GaAs, or $CuInSe_2$ because of their early rise in electronics and optoelectronic devices associated, in part, with the ease of doping them. Our general understanding of doping bottlenecks (i)-(v) above is primarily based on studying various would-be semiconductors. But rising interest in chemically different groups of compounds may now bring up different doping issues. Indeed, until recently, the condensed matter community used to refer to *interesting materials* as "highly correlated materials", i.e., those that presumably cannot be understood via the traditional mean-field band structure theory. This position is changing because of two realizations: (a) our condensed matter community found out that there is exciting quantum physics even in highly *un*correlated materials, all based on s-p electron bonding, such as topological insulators and semimetals illustrated by HgTe[35,36], $Bi_2Se_3$[37], or spin-orbit controlled Dresselhaus and Rashba materials illustrated by GeTe[38,39] and BiTeI[40]; (b) many of the interesting effects associated previously with explicit correlation effects in d- or f-electron chalcogenides and pnictides have now been addressed also by mean-field theory (provided that electronic and structural symmetry is allowed to break—see Sec. 2).[27,41-53] As the exclusive allure of "correlated materials" is diminishing, it has become more pragmatic to refer to interesting materials as "quantum materials" whether they are correlated or not. This changed material focus is no longer restricted to compounds containing "correlation agents" such as open-shell d- or f- electron elements but now has a different sort of structural and spin-orbit complexity that adds interesting doping challenges. The recent emergence of "quantum materials" including Mott insulators, new superconductors, spin-orbit dominated Rashba and Dresselhaus compounds, unusual magnets, qubit-enabling structures (e.g., hosting Majorana Fermions or point defect complexes in diamond), transparent conductors, topological insulators, Dirac and Weyl semimetals/metals, spin-liquids, and many more have quickly raised both practical and scientific questions regarding the possibilities of doping them.

*Current definition of Quantum Materials:* Definitions of quantum materials can be broad "…encompassing all materials whose properties are largely determined by quantum mechanical principles and phenomena. A key distinction of quantum materials from other materials lies in the manifestation of quantum mechanical effects at macroscopic length scales".[54] On the other hand, the Report of the Basic Energy Sciences Roundtable on Opportunities for Basic Research for Next-Generation Quantum Systems, 2017[55] is a bit more specific, regarding quantum materials as *"Assemblies of materials or arrangements of trapped ions or electrons in which the uniquely quantum interactions between the components are tuned to produce a specific behavior".* The report cited above states that implementing quantum information science *"remains a significant challenge for the scientific community",* …listing as some of the critical challenges *"how to allow quantum systems to interact and ultimately transfer information to their classical environments without interfering with their quantum coherence. This challenge requires a new understanding of what types of defects are allowed in these systems, and how these defects evolve during the synthesis and fabrication of Quantum systems, including the controlled introduction and manipulation of complex defect systems"*.

*Classes of Quantum Materials that encounter doping challenges vary. A few illustrative cases include:*

(i) Pristine insulators that would become *superconductors* if successfully doped, including $CsTlF_3$[56], $La_2CuO_4$[57], $Nd_2CuO_4$[58], or odd-parity topological superconductors such as $Sr_2RuO_4$[59].

(ii) *Topological insulators* such as in $Bi_2Se_3$[37] could be affected by spontaneous anion vacancy formation, a donor defect that renders them metallic. Also, would-be topological insulators such as $BaBiO_3$[33] require for its functionality a significant doping that will position the Fermi level higher in the conduction band.

(iii) *Doping of organic compounds* is generally achieved by the insertion of interstitial donors such as alkali atoms[8-10]; but can the currently limited doping menu of organic solids be expanded by including substitutional impurities?

(iv) Insulating building blocks such as $LaAlO_3$ and $SrTiO_3$ can lead to *conducting* $LaAlO_3/SrTiO_3$[60] interfaces; a leading mechanism[61] for this spontaneous effect is doping by surface point defects enabled by the polarity of $LaAlO_3$.

(v) Doping of *transparent conducting oxides*[62,63] (such as $CuAlO_2$[64], $Cr_2MnO_4$[65], or $In_2O_3$[66])**:** this functionality is achieved by starting from transparent, wide-gap insulators and instilling in them conductivity by extensive doping. Yet, most insulators cannot tolerate extensive doping without developing structural reorganization that causes carrier compensation. This might explain perhaps why the coexistence of transparency and conductivity is so rare.

(vi) *Hybrid organic-inorganic perovskites* $AMX_3$ (with A=organic molecule; M= Pb, Sn, and X=halogen) that have amazed the photovoltaic community by reaching an unexpected ~25% thin-film cell efficiency in just five short years of research.[67,68] Yet, doping of halide perovskites is notoriously difficult[69], so in the corresponding solar cell, one tries to avoid this step by creating instead interfaces with hole and electron transport media.

(vii**)** Doping of *quantum spin liquids* (*QSLs*)**.** The QSLs, such as kagomé Zn-Cu hydroxyl-halides, have been expected to offer insights into high-temperature superconductivity if doping would be possible. Yet, successful doping of QSLs seems elusive,[34] defeating thus far the dream of creating another paradigm for high-temperature superconductivity.

(viii) The "Half-Heusler" ABC compounds $A^{III}B^{X}C^{V}$ (e.g., ScPtSb), $A^{IV}B^{X}C^{IV}$ (e.g., ZrNiSn), $A^{IV}B^{IX}C^{V}$ (e.g., TiCoSb), and $A^{V}B^{IX}C^{IV}$ (e.g., TaIrGe) are emerging quantum materials[70,71] harboring topological, thermoelectric, and transparent conducting functionalities. Yet, their doping behavior is not understood, apparently, because built-in intrinsic non-stoichiometry obscures the effect of intentional doping.[72-74]

We do not aim to review all that has been done on doping quantum materials but focus on what we feel are the intriguing, if not peculiar aspects of the phenomenology. Indeed, this article reviews what are the top dozen unusual doping effects noted in the literature (Sec. 2), connecting them to concepts in the modern theory of doping (Sec. 4), and then discusses how these effects are explained by the underlying concepts developed in the modern quantum theory of doping (Sec. 5). In passing, Sec. 3 gives the essential background on the different historical views on what level of theory was needed to discern metals from dope-able insulators. This approach could forge a bridge between condensed matter theory of real materials vs. condensed matter theory of model Hamiltonians and the solid-state chemistry of doping.

## 2. DOPING PHENOMENOLOGY AND DOPING PECULIARITIES

Whereas the previous section focused on the new quantum *materials* that require doping for achieving their specific functionalities, the present section deals with the new *phenomena* observed in doped quantum materials that any successful theory of doping might aim to explain.

*The self-regulating characteristic of doping:* One of the confusing aspects of doping quantum materials is that the host crystal is not always "electronically rigid" with respect to doping, but instead can rearrange structurally in a Le Chattelier-like manner in response to the introduction of carriers. Here, we distinguish the intentional doping by impurities from the process occurring in the background as the feedback response of the solid to such doping. Indeed, structural defects such as cation vacancies (often hole producing acceptors) or anion vacancies (often electron producing donors) can form spontaneously as a result of deliberate n-type or p-type impurity doping, respectively.[24-26] This doping-induced formation of selective structural defects acts as a limiting factor, inhibiting further deliberate doping. Compounds susceptible to the formation of anion (cation) vacancies will thus experience intrinsic limitations to deliberate n-type (p-type) doping. Whereas textbooks[75-78] often assume that materials can be doped to achieve practically any desired target position of the Fermi level within the band gap, it is now becoming apparent that the Fermi level in compounds has a mind of its own. For example, inserting into some compound impurities believed to be soluble and readily ionizable electron donors do not always end up shifting the Fermi level, causing instead "Fermi level pinning". In fact, the energy window available for shifting the Fermi level in some materials might be frustratingly narrow. The understanding of Fermi level pinning and the difficulties to dope to a target Fermi level. This will be explained in Sec. 5.1.

*Compounds could exhibit intrinsic asymmetry between doping electrons vs. doping holes:* In the old days, it was believed that wide-gap insulators cannot be doped.[3] It now appears that many wide-gap insulators can indeed be doped, but only one way—either by electrons (n-type) or by holes (p-type). NiO, MgO, and ZnO provide good examples, demonstrating a fundamental asymmetry[79,80] in introducing free carriers into a solid. Specifically, ZnO can be readily doped by electrons but not by holes, whereas NiO is readily dopable by holes but not by electrons, and MgO can hardly be doped by either. In semiconductors, most II-VI *Telluride* compounds and most III-V *Antimonides* compounds are predominantly naturally p-type, and most III-V Arsenides compounds are predominantly n-type (see, for example, data compiled in Ref.[24]). This asymmetry is not affected by using different doping methods and thus seems to be an *intrinsic* tendency. The understanding of this asymmetry is provided in Sec. 5.2.

*Failing to dope to a target Fermi level can impede doping-induced topological insulation:* Band structure calculations of compounds that have band inversion at an energy $E^{inv}$ characterizing topological behavior are now rather common.[30-33] Unfortunately, many of these predictions show that band inversion occurs high in the conduction band or deep inside the valence band, thus requiring significant doping to reach $E^{inv}$ and realizing such exciting topological functionality. For example, it was proposed that certain compounds, e.g., BaBiO$_3$ could become topological insulators by doping[33], creating a new class of doping-induced topological insulators. Missing is a structural theory of doping that predicts how far the Fermi level can be shifted without creating "collateral damage", transforming the material to another structure that might not necessarily be topological.[81-83] This subject will be discussed in Sec. 5.3.

*Doping rules and designer dopants:* The field of practical doping is full of anecdotal observations[24,26,84-92] regarding which compounds can be doped readily p-type (e.g., ZnTe, CdTe, InSb, and GaSb) and which can be doped readily n-type (e.g., InP, InAs, ZnO, and CdSe) as well as advice given as to how to dope insulating perovskites and spinels p-type or n-type. One also often encounters "non-intuitive dopants": whereas textbook's view of *acceptor* dopants creating holes refers to an impurity having lower valence than the host atom being replaced (such as B in Si), whereas donor dopants creating electrons must have higher valence than the host atom being replaced (such a P in Si), surprisingly, some nominally isovalent impurities such as Rh replacing Ir in Sr$_2$IrO$_4$ act as acceptors[93]. We will explore in Sec. 5.4 what are the pertinent physical "design principles" that can be distilled from theory to explain this phenomenology.

*Doping by natural off-stoichiometry:* Daltonian view[94] of fixed, integer stoichiometry in compounds at low temperatures has been a cornerstone of inorganic chemistry. Yet, systematic deviations from the fixed Daltonian integer ratios show up even at low temperatures. For example, ABC half-Heusler compounds with a 3d element in the B-site tend to systematically be B-rich[73], whereas compounds with B = 5d elements (e.g., ZrPtSn), according to limited experimental data[95], tend to exhibit an imbalance in the stoichiometric ratio between the C and A elements. Such deviations from the fixed Daltonian integer ratios seem to correlate with doping: the compounds containing 3d elements in the B position (ZrNiSn and ZrCoSb) turn out to be naturally n-type[73], whereas the compounds containing 5d elements in the B position (ZrPtSn and ZrIrSb) are often p-type.[95] The understanding of the relationship between intrinsic trends in non-Daltonian stoichiometry and doping trends is in its infancy and will be discussed in Sec. 5.5.

*Spontaneous defect formation can remove exotic topological features:* Some quantum materials can exhibit the spontaneous formation of point defects that destroy their unique topological properties. For instance, $Ba_4Bi_3$ was expected to have special quasiparticle symmetry associated with its presumed metallic band structure having the Fermi level within the principle valence band (i.e., being a hole-rich metal).[29] This electronic configuration was later shown to be unstable towards the spontaneous formation of Bi vacancies[82] that self-doped the system by electrons, thus compensating the holes while also breaking the symmetry that promised the exotic topological properties. This will be explained in Sec. 5.6.

*The coexistence of transparency and conductivity in some oxides as a special form of ultra-doping:* Most transparent substances are insulators, whereas most metallic conductors are opaque. Transparent conductors seem to violate this contraindication where in rare cases, optical transparency coexists with metallic conductivity, e.g., p-type $CuAlO_2$[64] and $Cr_2MnO_4$[65] or n-type $In_2O_3$[66]. The understanding of the factors controlling the coexistence of conductivity and transparency as an extreme form of doping will be discussed in Sec. 5.7.

*Atomic defects vs. electronic defects:* The traditional view of doping is *atomistic,* such as the doping by atomic impurities substituting a host atomic site. The deciding quantity is the *atomic perturbation* $\Delta V_{imp}=V(I)-V(H)$ between the potential of *the substitutional* impurity atom (I) and the host site (H). When $\Delta V_{imp}$ exceeds a threshold value, an impurity level would split from the continuum valence or conduction bands into the band gap region. This atomistic picture of Koster and Slater[96] and its extension to 3D semiconductors by Hjalmarson et al.[97] directed the field of impurity doping to consider atomistic constructs such as I vs. H atomic electronegativity or size mismatch, etc. However, doping can also occur as a result of an *electronic perturbation* forming levels inside the band gap without intervention of a substitutional impurity that is integrated into the lattice structure. This can be achieved by photodoping (removal of a carrier by its photoionization), modulation doping (spatially separating the ionized dopant from the free carrier), gating, or creating a weakly coupled *interstitial* impurity (Li, H) that gets ionized without being integrated into the lattice structure as substitutional impurities do. Such "*electronic impurity*" can involve a $Ti^{4+}(d^0)$ ion in a solid that captures an electron and reconstructs it to $Ti^{3+}(d^1)$. The perturbation potential driving the formation of such an electronic defect in this example is then $\Delta V_{elec}= V(Ti^{3+}) - V(Ti^{4+})$, replacing the Koster-Slater atomic perturbation $\Delta V_{imp}=V(I)-V(H)$. The two leading categories of electronic impurities include (i) electron or hole *polarons* (e.g., electron polaron in e-doped $TiO_2$[98-100] or hole polaron in h-doped $SrTiO_3$ [6,201]) whereby a localized state is formed in the gap such as $Ti^{3+}$ state with wavefunction localized on *single* reduced cation Ti site, and (ii) formation of split-off in gap electron or hole intermediate <u>bands</u> such as in $YTiO_3$, $LaTiO_3$, and $YNiO_3$. The two forms (i) and (ii) of electronic perturbations leading to trapped carriers describable by density functional theory (DFT) will be described in Sec. 5.8.

*Doping by polarons and the percolation problem:* The story of high-$T_c$ superconductors is largely the story of doping-induced superconductivity. Large classes of superconductors start as insulators and acquire their functionality only once the Fermi level is displaced from the insulating band gap region into the valence or conduction bands.[101] An interesting electronic doping scenario is the capture of a carrier that leads to the formation of *localized polarons or charge density waves* that normally does not result in high conductivity. However, when the concentration of such polarons increases to the point that they percolate, creating a wall-to-wall continuum, such polarons lead to real doping and the ensuing conductivity. Also, creation of vacancies in simple oxides such as CaO or $HfO_2$ was once proposed to be a source of itinerant ferromagnetism (without the presence of magnetic ion[102-108]). However, the concentration of vacancies needed to create a percolating wall-to-wall network of moments calculated in Ref.[109,110] is unrealistically large, suggesting that ferromagnetism in such nonmagnetic oxides may be hard to realize without many orders of magnitude deviation from equilibrium defect concentration. This idea of doping by the percolation of polarons or local moments will be explained in Sec. 5.9.

*Antidoping behavior: When the Fermi level shifts in the wrong way:* In systems that already have a polaron or a split-off intermediate band, one might wonder how would such electronically induced states behave if deliberately doped. Textbooks teach that n-type (p-type) doping shifts the Fermi level towards the conduction (valence) band. However, it has been recently noted experimentally[111-115] and theoretically[116-118] that *electron* doping of $Li_xFeSiO_4$[116], $SrCoO_{3-\delta}$[113], $Li_xIrO_3$[116], $YNiO_3$[118], and $SmNiO_3$[111,112,114-117] shifts the doped band into the principal *valence* band, leading to the increases in the band gap and thus *reduction of* the conductivity. This peculiar phenomenon has been named "antidoping" and will be discussed in Sec. 5.10.

*Doping can be altered fundamentally by the polarity of the medium:* The undoped interface between nonpolar $SrTiO_3$ and polar $LaAlO_3$ is known to show two-dimensional electron gas (2DEG) above a critical $LaAlO_3$ thickness[60], even though the two corresponding bulk compounds are wide-gap *insulators*. This was attributed to the "polarity catastrophe" scenario[119], whereby an ideal, defect-free crystal would lead to ionization of the electrons from the valence band of $LaAlO_3$, transferring them to the $SrTiO_3$ interface. However, this requires the overlap of the valence band maximum (VBM) of $LaAlO_3$ with conduction band minimum (CBM) of $SrTiO_3$, contrary to the experimentally observed negligible band-bending[120-123] in the $LaAlO_3$ film, indicating that the $LaAlO_3$ VBM is located energetically far below the Fermi level. There is increasing evidence that the surprising interfacial conductivity is caused by polarity that reduces the formation energy of carrier-producing surface defects[61], being then swept to the interface. This will be described in Sec. 5.11.

*Interstitial hydrogen as a common dopant for organic and inorganic compounds and the emergence of universal doping:* A common practice in otherwise rather different traditions in doping organic vs. inorganic insulators is the use of interstitial monovalent cation dopants. Organic doping is often done by alkali metals[8-10], whereas the inorganic analog is interstitial hydrogen.[11,15,16] Calculations by Van de Walle et al.[13] indicated that, in most cases, interstitial hydrogen exists in three charge states $H^+$, $H^0$, and $H^-$ forming a single, two-electron transition level $\varepsilon(+/-)$. The interesting observation[12] is that hydrogen insertion renders some oxides metallic, yet others stay insulating upon hydrogen doping. These observations led to the interesting conclusion that the $\varepsilon(+/-)$ transition level of hydrogen forms a global demarcation energy that predicts conducting vs. insulating behaviors. This will be expanded upon in Sec. 5.12.

## 3. DETERMINING IF AN UNDOPED COMPOUND IS A METAL OR INSULATOR: THE NEEDED THEORETICAL FRAMEWORK

**3.1 What is the minimal theoretical level needed to explain gapping, hence dopability of d-electron insulators:** Doping of a compound starts by recognizing if, in its undoped form, it is a metal or insulator. The experimental distinction is straightforward, based on measuring temperature-dependent conductivity. As far as theory is concerned, however, numerous insulators were misconstrued using simplified forms of DFT band structure to be metallic[53] and thus were not considered as viable candidates for the theory of doping. For example, assuming for Rocksalt CoO or NiO, a minimal unit cell (with a single formula unit per cell) forces the equivalence of all-metal sites in the lattice, justifying a nonmagnetic configuration that predicts a metallic state for an odd number of electrons/formula unit. The absence of long-range order in such nonmagnetic unit cells leads to a metallic band structure. This can be avoided if different spins are allowed to "see" different potentials, leading to gap formation in the spin-polarized description of Slater antiferromagnets (AFM) even in mean-field DFT band theory.[124] However, whereas long-range order of spins in antiferromagnets solved the Mott dilemma, the paramagnetic (PM) phase lacking such order was viewed to still be described incorrectly by band theory as metals. Indeed, if the paramagnetic phase is described as a nonmagnetic configuration, the mean-field theory will produce a metallic structure again, in conflict with many experiments. A recent review[53] summarizes many cases in oxide physics where a simple version of DFT incorrectly produced false metals.

The question that arose was what causes the failure of band theory in predicting incorrectly metallic states in a large number of known insulators. One way to introduce gaps in band structure metals is via the Mott-Hubbard mechanism. It uses the picture that the occupation of lattice sites by electrons or the avoidance of such occupation is a result of electron-electron repulsion (codified by the Hubbard U interelectronic term). In this view, the formation of band gaps in such quantum materials, as needed for eventual doping, would require specialized correlated methodologies that go beyond what simple DFT band theory would accommodate.[125,126]. Consequently, d-electron compounds such as NiO, MnO, or $ABO_3$ perovskites, where B is a d-electron atom, became known almost synonymously as "correlated materials". The failures of naïve (N) DFT to predict the experimentally observed insulating states of numerous 3d oxides were interpreted in the correlated literature as evidence for the crucial role of electron correlation. For example, in the case of rare-earth nickelates $RNiO_3$, it was stated[127] that "*Standard DFT and DFT+U methods fail to describe the phase diagram, with DFT predicting that all compounds remain metallic and undisproportionated…. These results establish that strong electronic correlations are crucial to structural phase stability and methods beyond DFT and DFT+U are required to properly describe them*". DMFT study[128] of $La_2CuO_4$ motivated the need for correlation by stating "*While density functional band theory is the workhorse of materials science, it does not capture the physics of the Mott/charge-transfer insulator transition*". DFT-DMFT study[129] on NiO explained "*Conventional band theories which stress the delocalized nature of electrons cannot explain the large gap and predict NiO to be metallic. For this reason, NiO has long been viewed as a prototype "Mott insulator"*". Ref. [130] noted about the inability of LDA, GGA, and LDA+U to describe atomic distortions in 3d oxides "*However, these methods usually fail to describe the correct electronic and structural properties of electronically correlated paramagnetic materials*", "*Therefore, LDA+U cannot describe the properties of $LaMnO_3$ at $T > T_N$ and, in particular, at room temperature, where $LaMnO_3$ is a correlated paramagnetic insulator with a robust JT distortion*", "*LDA + U cannot explain the properties at $T>T_N$ and, in particular, at room temperature, where $KCuF_3$ is a correlated paramagnetic insulator*", and "*The nonmagnetic GGA calculations not only give a metallic solution but its total-energy profile is seen to be almost constant… This would imply that $KCuF_3$ has no JT distortion…, which is in clear contradiction to experiment*". Indeed, it seems that the correlated literature defines correlation as "everything that DFT does not get right" and that many effects (gaping paramagnets, orbital order, mass enhancement, atomic distortions, and disproportionation) were advertised as being "correlation induced".

One might wonder, indeed, if it matters how one refers to the effects that open a gap in false metals: dynamic correlation or mean-field DFT. We believe that it does matter because what is at stake is what is the physical picture/mechanism underlying such effects: the intrinsic mutual avoidance of electrons described by quantities such as pair correlation functions G(r,r') and on-site Coulomb repulsion U, or the symmetry breaking mean-field DFT theory? This distinction also determines what is the proper physical guide to be used in designing d-electron materials with target properties. This section discusses calculations that have misconstrued insulators for (false) metals, whereas proper (not highly simplified) DFT predicts instead insulating behavior. Thus, leapfrogging from simple band theory to correlated theories is not forced upon us by the failure of N-DFT.

**3.2 Naïve DFT often predicts false metallic states in numerous quantum materials:** The "False Metal Syndrome" abounds even in recent applications of DFT, based on the least number of possible magnetic, orbital, and structural degrees of freedom. The main restrictions used in such N-DFT involved (i) placing spatial constraints on unit cell symmetry and cell sizes (e.g., using a minimal/primitive unit cell of the highest symmetry); (ii) not allowing symmetry breaking, energy-lowering modes: this can miss atomic displacements, octahedral tilting and rotation, Jahn Teller (JT)[131] modes, bond disproportionation into small vs. large octahedra, all of which can lower the DFT total energy while removing degeneracies, opening the gap; (iii) applying constraints on the spin configuration in a para phase: often N-DFT (reviewed in Ref.[53]) uses a nonmagnetic view of paramagnets or non-polar views of paraelectrics (reviewed in Ref.[132]), considering that the global zero moment definition of a para phase can be interpreted on an atom-by-atom basis. This picture is guaranteed to produce a (false) metallic state for systems with an odd number of electrons/formula unit, creating the original Mott conundrum that required enlisting correlated methods to resolve; (iv) restrictions on the exchange-correlation (XC) functional used: N-DFT often used

(a) XC functionals (e.g., Perdew-Burke-Ernzerhof (PBE)[133] or PBEsol[134]) that fail to distinguish occupied from unoccupied orbitals and (b) has insufficient cancellation of self-interaction error as monitored by the failure of the total energy vs. occupation number to be linear (Koopman's compliance[135]). Here, (a) often leads to a preference for metals, and (b) to overly delocalized orbitals that may not have the needed spatial resolution to "see" symmetry breaking (i.e., they are "short-sighted").

The point is that the N-DFT methodology is not all that actual DFT can do. Indeed, there are avenues for removing the constraints on N-DFT other than disposing of DFT altogether, leading to correct gap formation in terms of structural, orbital, and magnetic symmetry breaking, all sanctioned by the DFT single-determinant, mean-field framework, without appeal to dynamic correlation. Other groups, including Sun et al.[47] and Ghosez et al.[48-50], have also recently used symmetry-broken DFT for the usual-suspect correlated compounds. The following section illustrates the symmetry-breaking mechanisms sanctioned by DFT band structure that lead to an insulating state (Fig. 1b) even though N-DFT (Fig. 1a) would falsely predict a metal. *This sets the stage for the DFT-based theory of doping all insulators.*

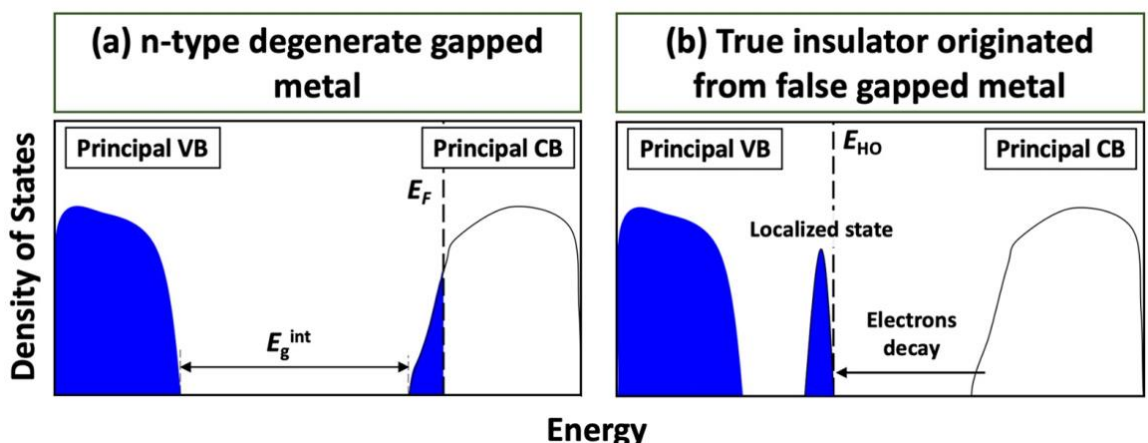

**Figure 1.** (a) Schematic density of states for degenerate gapped metal having Fermi level in the principal conduction band and an "internal" band gap ($E_g^{int}$) between the principal valence and conduction bands. (b) Density of states for a real insulator originated from a false metal where spontaneous symmetry breaking (e.g., structural relaxation, magnetization, or vacancy formation) resulted in band gap opening due to electron localization. The figure is reprinted with permission from Ref. [118], Copyright (2020) by the American Physical Society.

**3.3 Symmetry-broken DFT distinguishes false metals from dopable insulators and explains other effects previously attributed exclusively to correlation:** The symmetry-broken DFT approach relies on a few key factors ignored in N-DFT calculations: (i) the use of an XC functional that distinguishes occupied from unoccupied orbitals and renders orbitals spatially compact; (ii) the use of larger-than-minimal unit cells that do not force artificially high symmetry; (iii) "nudging" of the system to allow both local displacements and to break the orbital occupation pattern; (iv) possibility for the unequal occupation of previously degenerate states (e.g., for doubly degenerate E level with occupations E(x,y), we use $E(1,0)$ instead of $E(\frac{1}{2};\frac{1}{2})$]. Accounting for symmetry breakings allows the existence of a distribution of different structural or spin local environments and allows to explain gaping and orbital ordering[41-45], revealing quantitative agreement with measured local moments[41,43] and atomic displacements[45].

An often-overlooked key characteristic of proper DFT is the existence of a generally nonlinear feedback response connecting the electronic structure with the descriptors of the physical system: the Atomic identities, Composition, and Structure (ACS). Many features of the physical structure are determined iteratively by total energy minimization when seeking variational solutions. Thus, atomic displacements, octahedral tilting, charge redistribution and hybridization, and different spin arrangements capable of removing degeneracies are explored, often on-the-fly. In turn, the resulting energy lowering local symmetry breaking can affect the wavefunction. It is this nonlinear feedback, characteristic of *locally responsive electronic structure method*s that distinguishes such approaches from those model Hamiltonian methods that (i) have fixed physical descriptors leading but to a single possible outcome, (ii) restrict the explorable structural or magnetic degrees of freedom by using high symmetry structures (e.g., rather than supercells), and (iii) may be "short-sighted" when overly delocalized orbitals are used, lacking the needed spatial resolution to "see" and couple to local symmetry breaking. This feedback loop between structure and electronic properties characterizing responsive electronic structure methods (even at the mean-field level) is at the heart of understanding many of the peculiarities of quantum materials and their doping. In what follows, we illustrate (Fig. 2) how systems that were previously misconstrued by simple theory to be metals because of correlation deficiency are now recognized by symmetry-broken mean-field DFT to be insulators, thus amenable to doping.

| Cause of False Metal | Cause of gapping false metal | Example |
|---|---|---|
| Ignoring local magnetic motifs | Magnetic order | $CuBi_2O_4$, NiO |
| | Local spin environment | Paramagnets (e.g., $LaTiO_3$, $YTiO_3$) |
| Ignoring local structural/orbital motifs | Atomic displacement | $SrBiO_3$, $YNiO_3$ |
| | Octahedral tilting | |
| | Disproportionation | |
| | $Q_2^+$ distortion | $LaMnO_3$ |
| Ignoring defect induced symmetry breaking | Defect formation | $Ba_4As_3$, $Ag_3Al_{22}O_{34}$ |
| Ignoring spin-orbit coupling | Allowing spin-orbit coupling | $CaIrO_3$, $Sr_2IrO_4$ |

**Figure 2.** False metals and set of symmetry-breaking energy lowering mechanisms resulting in band gap opening with illustrative examples.

**3.3.1 Polymorphous distribution of local spin environments can convert a nonmagnetic false metal to a real paramagnetic insulator:** Until recently, the properties of paramagnetic phases have been explored as properties of *average nonmagnetic* structures <$S_0$>.[129,136-139] In this approximation, PM phases of Mott insulators were metallic, very often in contradiction with the experiment. For instance, nonmagnetic calculation fails to predict insulating disproportionation in $YNiO_3$ resulting instead in a metallic state for orthorhombic $YNiO_3$ (space group number (SG): 62) as shown in Fig. 3a. However, experimentally, a low-temperature phase of $YNiO_3$ is monoclinic (SG: 14) insulator having distinct structural disproportionation.[140,141] It has been demonstrated recently that modeling the paramagnetic compound as a spin-disordered system (each atom has non-zero spin, but the spins in the system are not locally ordered) can be used to effectively reproduce experimental properties of binary[41] and ternary[42,43] PM compounds within DFT calculations. The application of this PM model to $YNiO_3$ (SG: 14) results in an insulator (Fig. 3b). Importantly, the electronic structure of low-temperature paramagnetic $YNiO_3$ resembles that of low-temperature antiferromagnetic $YNiO_3$ phase, which is also an insulator with band gap energy of 0.59 eV (Fig. 3c). The analysis of magnetic moments for different spin configurations (Fig. 3) suggests that while in nonmagnetic $YNiO_3$ all spins are compensated locally, both AFM and PM configurations have a distribution of local non-zero magnetic moments, suggesting indirectly that PM-AFM phase transition can be considered as stabilization of certain spin and structural motifs existing in PM structure. These results thus demonstrate that

one can describe sufficiently well the electronic properties of paramagnetic compounds as long as they are described as average $P_{obs}=\Sigma P(S_i)$ of the properties $\{P(S_i)\}$ of the individual, low symmetry microscopic configurations and not the properties of globally average structures.

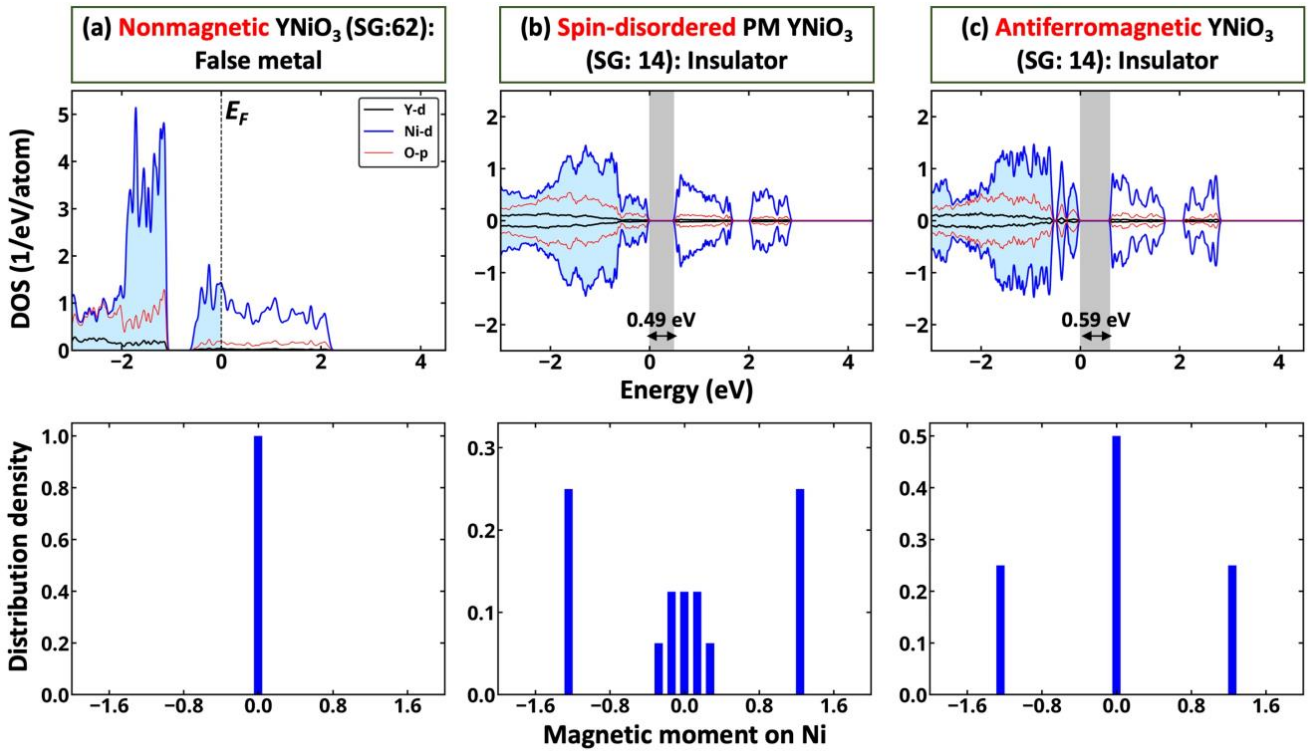

**Figure 3.** Electronic density of states [(a)-(c) top row] and the corresponding distribution of local magnetic moments [(a)-(c) bottom row] for nonmagnetic, paramagnetic (PM), and antiferromagnetic YNiO$_3$. (a) Naïve nonmagnetic model described in a primitive cell results in a metal state for YNiO$_3$ (SG: 62) with all Ni atoms having zero magnetic moments. (b) Paramagnetic YNiO$_3$ (SG: 14) described in a supercell with all Ni atoms having non-zero magnetic moments adding up to zero total magnetic moment is an insulator with the electronic properties and the distribution of magnetic movements resembling those of the low-temperature antiferromagnetic phase shown in (c). (c) Antiferromagnetic YNiO$_3$ is an insulator in the disproportionated cell (SG: 14) with half of Ni atoms having zero magnetic moments. Occupied states are shadowed in light blue. The band gap is shown in gray. SG denotes the space group number. The figure is drawn using data from Ref.[53].

**3.3.2 Energy-lowering bond disproportionation can convert a false metal into an insulator:** In simplified band calculations, pristine compounds are usually described with the smallest possible primitive cell, where each species is often represented via a so-called "single local environment" (SLE), as shown in the inset of Fig. 4a. While this minimal unit cell approach can be used for a range of compounds, for cubic SrBiO$_3$ (SG: 221), such model results in a (false) metal electronic structure with the Fermi level in the valence band and an internal gap above it (Fig. 4a). However, SrBiO$_3$ is an insulator.[142] Fig. 4b illustrates that the allowance of energy lowering bond disproportionation in SrBiO$_3$(SG: 221) results in band gap opening. Supercell of SrBiO$_3$(SG: 221) spontaneously disproportionates to monoclinic SrBiO$_3$ (SG: 14)[142], where Bi has a double local environment. The resulting system is an insulator containing symmetry inequivalent Bi atoms with distinct average Bi-O bond lengths of 2.17 and 2.34 Å as explained in Ref. [53].

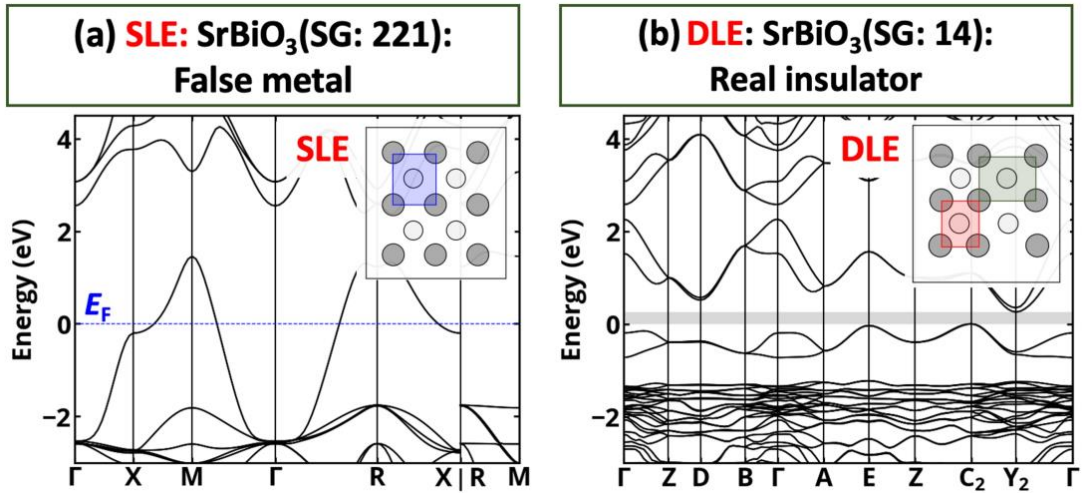

**Figure 4.** Structural symmetry lowering (transition from the single local environment (SLE) to double local environment (DLE)) results in a band gap opening for cubic SrBiO$_3$. (a) Band structure for SLE cubic SrBiO$_3$ (SG: 221) showing that the compound is a degenerate gapped metal with the Fermi level in the principal valence band. (b) Band structure for lower energy DLE monoclinic SrBiO$_3$ (SG: 14) showing that the compound is an insulator with band gap energy (shown in gray) of 0.26 eV. The insets schematically demonstrate the SLE and DLE structures. SG denotes the space group number. The figure is redrawn using data from Ref. [53].

**3.3.3 Gapping due to allowance of Jahn-Teller-like distortion:** In simplified DFT calculations, the properties of perovskites are often described using the structures with disallowed octahedra distortions, i.e., high-symmetry cubic phases. For quantum materials (i.e., Mott insulators), such model can predict metallic electronic structure as demonstrated for AFM LaMnO$_3$ (Fig. 5a). However, the low-temperature phase of LaMnO$_3$ is a wide band gap insulator with clearly distinct octahedra distortion.[143,144] Allowing structural symmetry breaking via using a large cell with structural nudging results in energy lowering, band gap opening, and clearly district octahedra distortion pattern (Fig. 5b). The band gap opening mechanism has been recently explained by Varignon et al.[46], who demonstrated that the ground state LaMnO$_3$ insulator structure has pseudo-Jahn-Teller distortion – the $Q_2^+$ octahedral deformation is a consequence of pure semiclassical atomic size effects resulting in octahedral rotations/tilts and not induced by an electronic instability, which is the origin of true Jahn-Teller distortion.

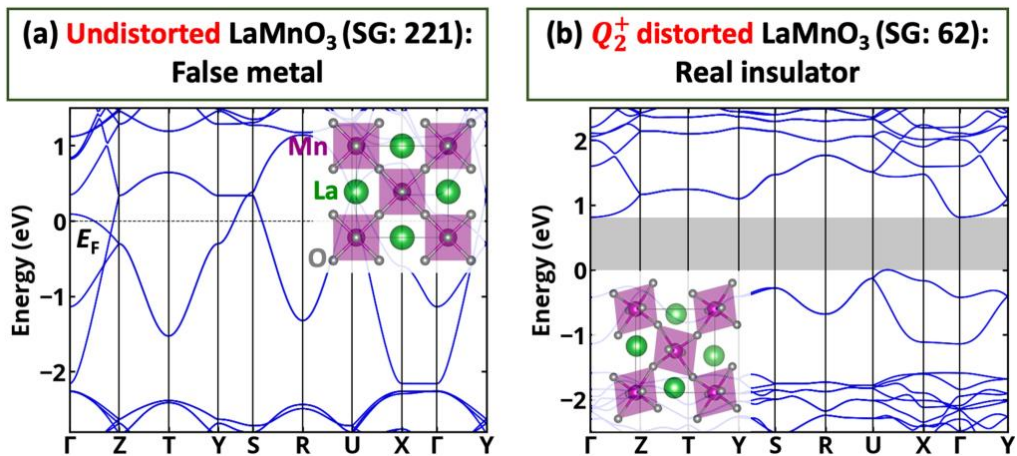

**Figure 5.** (a) Ignoring the pseudo-Jahn Teller $Q_2^+$ distortion, one predicts a metal in AFM LaMnO$_3$; (b) allowing energy lowering $Q_2^+$ distortion results in an insulator. The insets depict the corresponding octahedra patterns. The band gap is shown in gray. SG denotes the space group number. The figure is redrawn using data from Ref. [53].

**3.3.4 Using soft exchange-correlation functional can lead to false metals:** Certain XC functionals (e.g., PBE) produce insufficiently compact orbitals or orbitals that do not cancel sufficiently the self-interaction error, often predicting false metallicity. This is evident in part in modern databases of electronic structures (e.g., Materials Project[145], OQMD[146], and AFLOW[147]) where calculations are generally done mainly using soft XC functional such as PBE that has significant "delocalization error"[148] and cannot capture the formation of the in-gap state (Fig. 1b) that results in band gap opening.[53,100,139,149-152] However, the insulting nature of these and many other compounds can be captured with XC functionals that are known to have smaller delocalization error even without an artificial "+U" parameter (e.g., SCAN[153] or hybrid functional[154,155]).

**3.3.5 Spontaneous defect formation in a defect-free false metal can convert it into an insulating ordered vacancy compound:** Degenerate gapped metals illustrated in Fig. 1a can develop an instability with respect to the formation of intrinsic defects. Indeed, while for typical insulators, the formation of a point defect is limited to high temperatures, for degenerate gapped metals with an internal band gap (Fig. 1a), a split-off state can readily form in the gap (Fig. 1b) so that the carriers in the conduction band decay into such a band. Such energy lowering due to electron-hole recombination can overcome the energy need to create the point defects as has been recently demonstrated for n-type degenerate gapped $Ag_3Al_{22}O_{34}$[156] metal, where the acceptor Ag vacancies from spontaneously.[157] The same phenomenon has been shown for a potential p-type degenerate gapped $Ba_4As_3$ metal, demonstrating the spontaneous formation of As vacancies.[53] In the concentrated limit, such vacancies can result in non-stoichiometric systems explaining why attempts to stabilize $Ba_4As_3$ resulted in As-poor insulating $Ba_4As_{3-x}$ compound.[158]

**3.3.6 Allowing for spin-orbit coupling in a metal can convert it into a real insulator:** Traditional first-principle explorations of materials properties sometimes neglect spin-orbit coupling (SOC). However, for systems containing heavy elements, neglecting SOC can result in false predictions. For instance, $CaIrO_3$[159] and $Sr_2IrO_4$[160] were both predicted to be degenerate p-type gapped metal with the Fermi level in principal valence bands even when hard XC functionals (e.g., hybrid functionals) are used. However, experimentally both systems are insulators.[161,162] It turns out that description of band gap opening requires using both hard XC functional and including SOC.[159,160]

**3.3.7 In-gap multiplet states in wide-gap insulators:** The above subsections pointed out how allowance of various forms of energy-lowering symmetry breaking (Fig. 2) in mean-field DFT leads to a number of effects previously attributed exclusively to explicit dynamic correlation. Such effects include gapping of paramagnets[41-44,53], orbital ordering[42,43,49], mass enhancement[163], Jahn-Teller distortions[46], and nematicity[164]. We note, however, that single determinantal approaches such as standard DFT do not describe atomic-like multiplet effects. Such states can appear as d-d* transitions mostly inside the band gaps when d-electron impurities exist in wide-gap insulators.[165,166] For example, $Co^{2+}$ doping in cubic host crystals such as ZnSe with e and $t_2$ crystal orbitals can be described either in a $e^4t^3$ or in a $e^3t^4$ single-particle occupation, but in standard DFT, such configurations do not coexist at the same time, so they cannot interact. Such (and many more) single-particle configuration can create atomic-like multiplet states that, in reality, could interact and mix, leading to absorption and ionization transitions that differ significantly from their single-particle analogues. Examples are given in Fig. 6 for a variety of 3d ion impurities in wide-gap ZnSe insulator. Additional examples of 3d dopants and impurities in wide-gap insulators are provided in Ref. [165-168]. Such multiplet effects can be described theoretically in many ways[165-167] including by combining the classical ligand field impurity multiplet theory[166] with DFT calculation of the renormalization of the many-electron integrals in the solid relative to the free atoms, as described in Ref. [165,167]. Multiplet spectroscopic fingerprints of doping are apparent in particular where localized impurity levels (3d orbitals, or even the nitrogen-vacancy center in diamond[169-171]) in a wide-gap energy region and are thus isolated electronically from most level-broadening and level-coupling effects. Detailed description of these specialized cases of in-gap multiplet transitions in wide-gap systems is beyond the scope if the current article, as they generally do not contribute to doping.

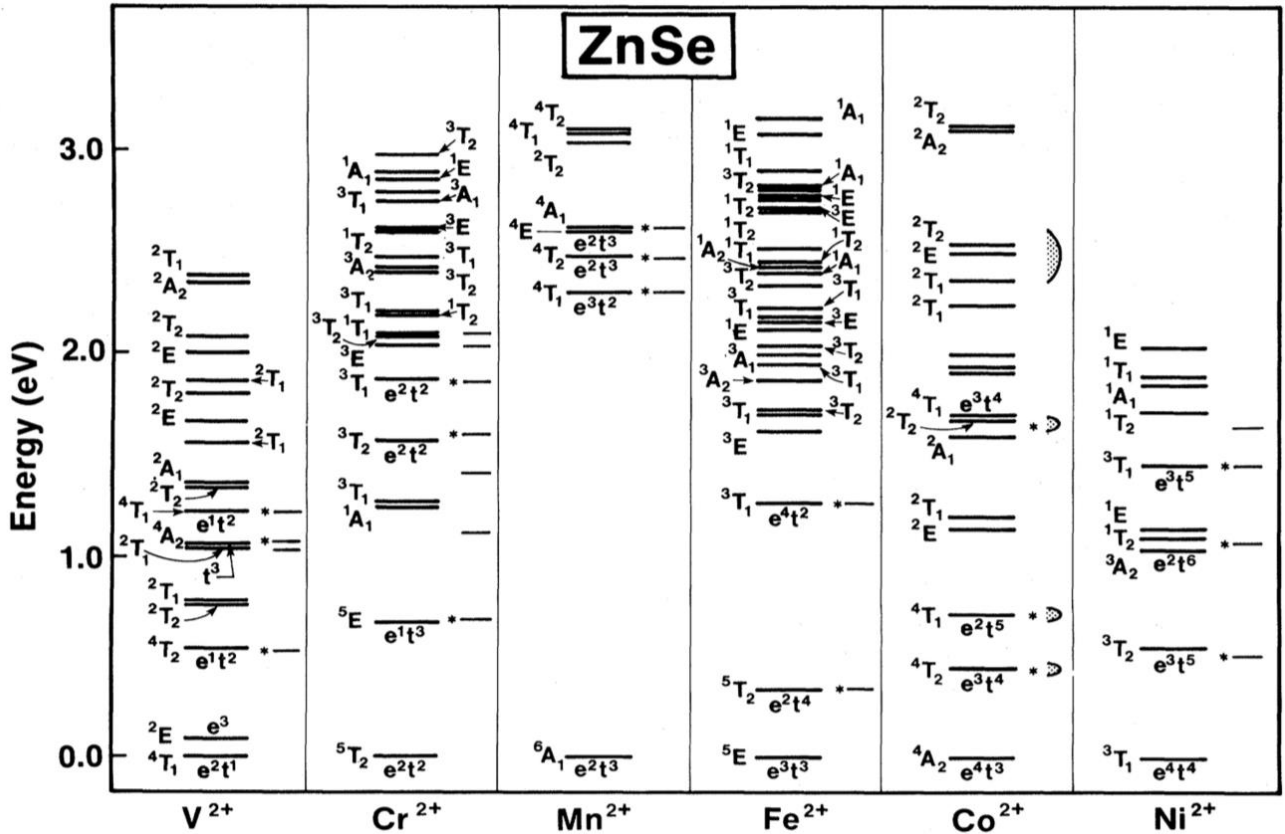

**Figure 6.** Calculated multiplet structure for 3d impurities inside the band gap of ZnSe. The experimental results (lines or bands represented as shaded areas) are indicated to the right of the calculated levels; asterisks denote the experimental transition energies used in the fit. The others are predicted. The dominant one-electron configurations as obtained in the present calculation are indicated for the main excitations. The figure is reprinted with permission from Ref. [167], Copyright (1984) by the American Physical Society.

Understanding the factors that led previously to theoretical misassignments of proper insulators as (false) metals described in subsection 3.3, now opens the way to the systematic investigation of doping in a wide range of quantum insulators. The density functional approach can achieve its results even without any U parameter (if the XC functional is close to Koopman's linearity).[43,44,47] If symmetry breaking is permitted to lower the total energy, DFT is no longer a "rigid band structure approach", but manifests instead a "responsiveness" of the electronic structure to different atomic positions and spin configurations that are explored via total energy minimization, creating a feedback response (an "elastic quality") of the electronic structure to the atomic and spin structures.

## 4. DOPING CONCEPTS UNDERLYING THE MODERN THEORY OF DOPING

Experience and insights gained by the semiconductor community in the past years in DFT studies of doping conventional semiconductors[172-181] can be leveraged, extended, and deepened in studying carriers in "Quantum Materials" such as d-electron oxides[87,182] or the nitrogen-vacancy center in diamond[169-171], once they are described by symmetry-broken DFT (Sec. 3). The basic approach involves a supercell containing the doping impurity or defect in equilibrium with a Fermi reservoir of carriers and a chemical reservoir of species that constitute the host compound as well as reaction products that could form between the impurity atom and the compound. The key concepts we wish to review in this section are somewhat different than those used in model Hamiltonian approaches of doping and include:

(1) Incorporation of dopants depends on the chemical potentials of the pertinent host and dopant species;

(2) The dopant formation energy is not a material constant but depends on the Fermi level, chemical potential, and charge state;

(3) Dopants are generally not charge neutral; their charge state decides the type of free carriers generated by doping;

(4) The dopant "transition level" is not an orbital energy but a difference in total energies and is given as the value of the Fermi level where the dopant switches its charge state. The transition energy does not depend on chemical potentials;

(5) The equilibrium thermodynamic Fermi level is determined by all charged species and decides the formation energy of each dopant and intrinsic defect;

(6) The predicted equilibrium concentration of dopants, defects, and free carriers is the final measure of doping success. The dependence of these quantities on external knobs, such as the parametric Fermi level and the chemical potentials of the constituents, is at the heart of understanding the trends and peculiarities in doping material reviewed in Sec. 2 and the key to optimizing them.

It is important to note that the culture in the computational field of doping 3d oxides and generally "correlated materials" has diverged significantly from the modern doping theory[172-181] based on the principles (1)-(6) above in that the critical dependences of the doping process on the factors discussed in subsections 4.1-4.6 below were generally not considered in the correlated literature.[183-186]

**4.1 Formation energy of dopant** D $\Delta H(D, q, \mu_\alpha, E_F)$ in a charge state *q* in equilibrium with a reservoir of elemental chemical potentials $\{\mu_\alpha\}$ and the Fermi sea with parametric energy $E_F$ (Fig. 7) is

$$\Delta H(D, q, \mu_\alpha, E_F) = (E(D,q) - E_H) \pm \sum_\alpha (\mu_\alpha^0 + \Delta\mu_\alpha) + q(E_V + E_F) + \delta H_{\text{corr}} \qquad (1)$$

where $E_V$ is the valence band maximum, $E_H$ is the energy of the undoped host compound, $E(D,q)$ is the energy of the supercell containing a dopant, $\mu_\alpha^0$ is a chemical potential of pure element, and $\Delta\mu_\alpha$ is excess chemical potential defined by synthesis conditions. The charge state q corresponds to donors (when q > 0) or acceptors (when q < 0). Formation energy of a donor $A^{+q}$ (q>0) increases as the parametric Fermi level moves from valence band maximum toward conduction band minimum, because the ionized donor electron must join the Fermi reservoir with energy $E_F$. In contrast, the formation energy of acceptor (q < 0) decreases for the corresponding Fermi level movements. However, for a given system, the Fermi level is not a free parameter and is determined by all charged defects present in the system (see Sec. 4.4). We note that the use of DFT to compute total energies might include uncertainties due to the imperfect exchange and correlation functional currently known, as evidenced in part by comparison with quantum Monte Carlo estimates.[187,188] Such differences call for corrections $\delta H_{\text{corr}}$ applied to DFT to account for supercell finite-size effects and systematic DFT errors in the band gap.[172-179] Furthermore, certain doped systems such as the nitrogen-vacancy center in diamond[170,171] or transition metal impurities in insulators[165,167] manifest atomic-like states in the gap that (just like free atoms) require many-body multiplet corrections, which can be accomplished by using DFT impurity wave functions in multiplet theory.

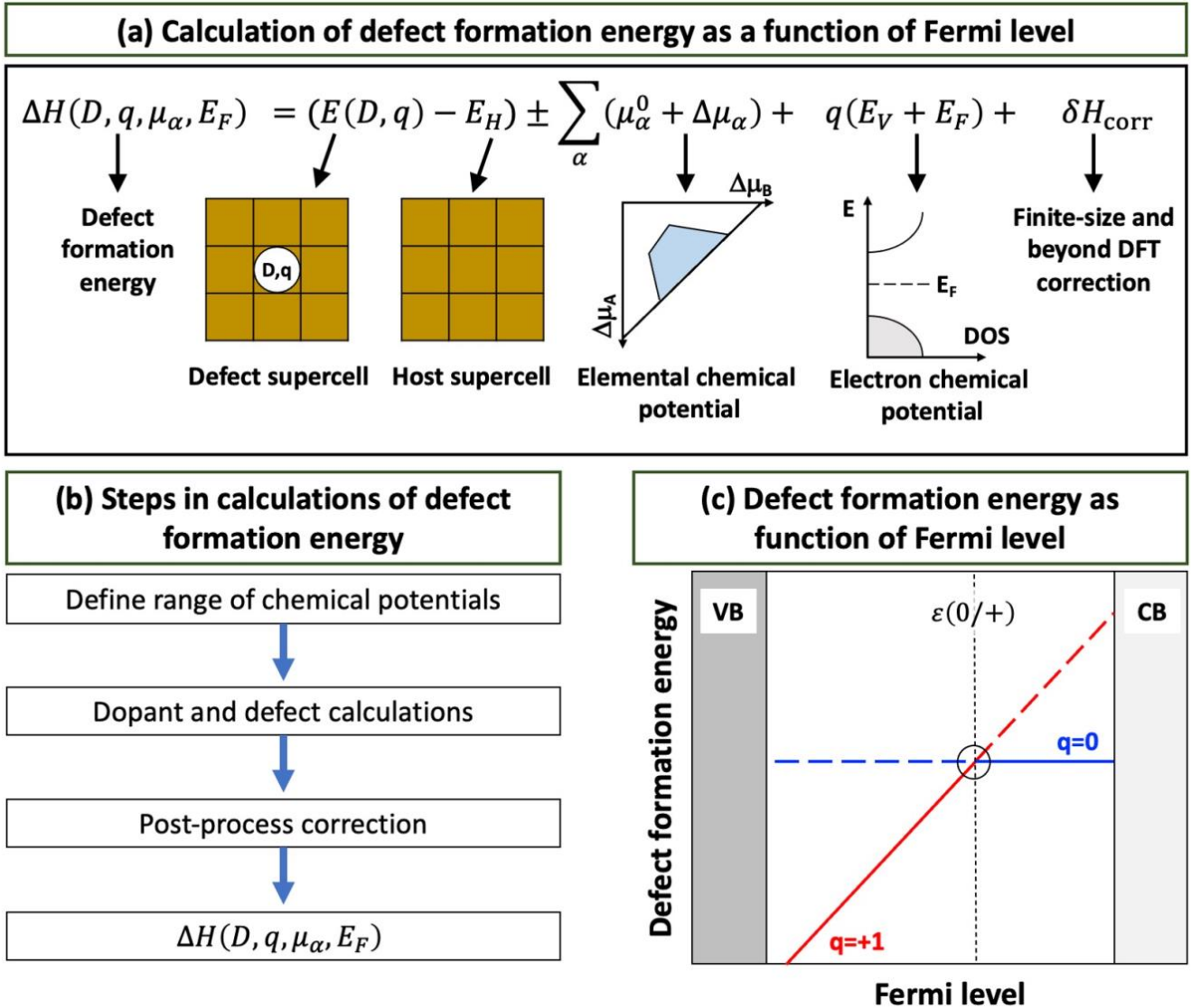

**Figure 7.** (a) Equation for defect formation energy as a function of the Fermi level and chemical potential. (b) Steps for calculations of defect formation energy. (c) Schematic illustration of the dependence of defect formation energy on the Fermi level and 0/+ transition level $\varepsilon(0/+)$.

**4.2 Charged dopant and charge transition level:** Eq. (1) indicates that dopant can exist in different charge states such as $A^+$, $A^-$, and $A^0$. Whereas the dopant formation energy of the charge-neutral defect $A^0$ is independent of the Fermi level, for charged defects $A^+$ and $A^-$, the defect formation energy depends on the Fermi level, and each defect can have the lowest formation energy at the specific range of $E_F$. Hence, one can define charge transition level $\varepsilon(D, q/q')$ of a dopant as the value of the Fermi level $E_F^*$, at which two charged states $q$ and $q'$ have equal formation energies $\Delta H(D, q, \mu_\alpha, E_F^*) = \Delta H(D, q', \mu_\alpha, E_F^*)$, which can be expressed as

$$\varepsilon(D, q/q') = \frac{(E(D,q) + \delta H_{\text{corr}}(D,q)) - (E(D,q') + \delta H_{\text{corr}}(D,q'))}{q' - q} - E_V \qquad (2)$$

We see that the transition energy $\varepsilon(D, q/q')$ is not an orbital energy or a quasi-particle energy but is defined by relaxed total energy of initial state q and relaxed total energy of final state q' appropriate to equilibrium measurements. To illustrate it, Fig. 7c displays a schematic of the dependence of dopant formation energy on Fermi level depicting that neutral dopant is the lowest energy defect near conduction band while the positively charged dopant has the lowest defect formation energy at the valence band; the charge transition level corresponds to the crossing point of the defect formation energies and is depicted as $\varepsilon(0/+)$.

**4.3 The diagram of allowed domains of chemical potential:** In order to perform doping calculations, one must first compute the range of chemical potentials corresponding to specific synthesis conditions under which the compound can exist instead of decomposing into any of its possible competing phases. For illustration, we consider ZrNiSn (B=3d) and ZrPtSn (B=5d) half-Heusler compounds and predict the range of excess chemical potential $\{\Delta\mu_A, \Delta\mu_B\}$ defined by $\Delta\mu_A + \Delta\mu_B + \Delta\mu_C \leq \Delta H$ (competing phases) and $\Delta\mu_{A,B,C} \leq 0$ under which the compounds are thermodynamically stable. By computing the formation energies of competing phases to ABC (such as $B_3C$, $A_2B_2C$, $A_5C_4$, and the full Heusler compound $AB_2C$), one predicts the range of chemical potentials under which ZrNiSn and ZrPtSn are stable as shown by green and yellow zones in Fig. 8. Importantly, that despite the similarity of the compounds, ZrNiSn and ZrPtSn exist in a different range of chemical potentials, illustrated by the different locations of the green and yellow zones within the stability triangle: ZrNiSn is stable with respect to decomposition *under B-rich* (A-poor) conditions as depicted in Fig. 8a, while ZrPtSn has a wider stability range of chemical potentials (Fig. 8b) and can be synthesized under both *A-rich (B-poor) and B-rich (A-poor) conditions.*

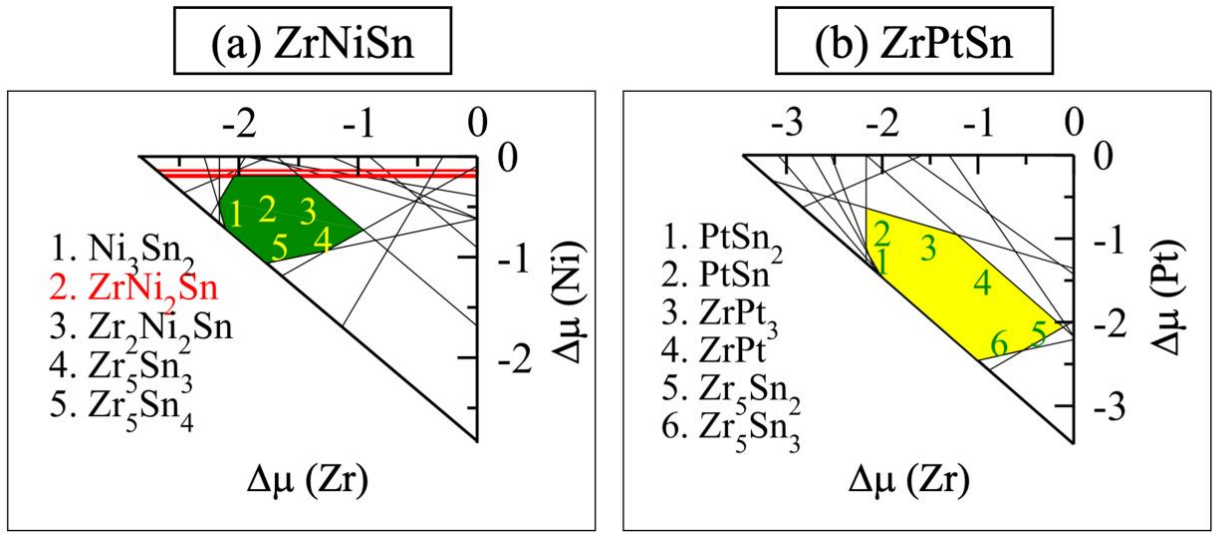

**Figure 8.** Comparison in the chemical stability field of (a) ZrNiSn (B=3d) and (b) ZrPtSn (B=5d) in terms of their excess chemical potential $\{\Delta\mu_A, \Delta\mu_B\}$ defined by $\Delta\mu_A + \Delta\mu_B + \Delta\mu_C \leq \Delta H$ (competing phases) and $\Delta\mu_{A,B,C} \leq 0$. The green and yellow areas are a guide to the eye for the dominance of A-rich and B-rich chemical potentials. The figure is reprinted with permission from Ref. [72], Copyright (2017) by the American Physical Society.

To gain insight into the critical dependencies in Eq. (1), Fig. 9 shows the calculated defect formation energies for ZrNiSn and ZrPtSn versus the parametric Fermi level for fixed chemical potentials defined in Fig. 8. We see that Ni interstitial in Fig. 9a is a donor (releases electron), shifting $E_F$ towards the conduction band, but the feedback response to shifting $E_F$ is the easier production (lower formation enthalpy) of Zr vacancies $V_{Zr}$, which are electron-killing acceptors. The effectiveness of the spontaneously created "killer defects" in a given doped ABC compound is the key in assessing its ultimate dopability. We see that the energy to create a Ni interstitial *increases* rapidly under Ni poor conditions, and the energy to create the Sn-on-Zr anti-site *increases* under Sn-poor conditions. These dependencies indicate the preferred growth conditions that one should select to maximize a target outcome in doping and do not depend on chemical potential. Fig. 9 also shows shallow acceptors such as Zr and Ni vacancies and Sn-on-Zr anti-site in ZrNiSn. If the donor level is inside the valence band $\varepsilon(+1/0) < 0$ or if the acceptor level is inside the conduction band $\varepsilon(0/-1) > E_g$, the corresponding defect does not create free carriers. A defect that has a "*shallow level*" $\varepsilon(D, q|q')$ close to the band edge can be ionized, but if its formation enthalpy $\Delta H(D, q, \mu_\alpha, E_F)$ is large, the concentration of carriers is negligible.

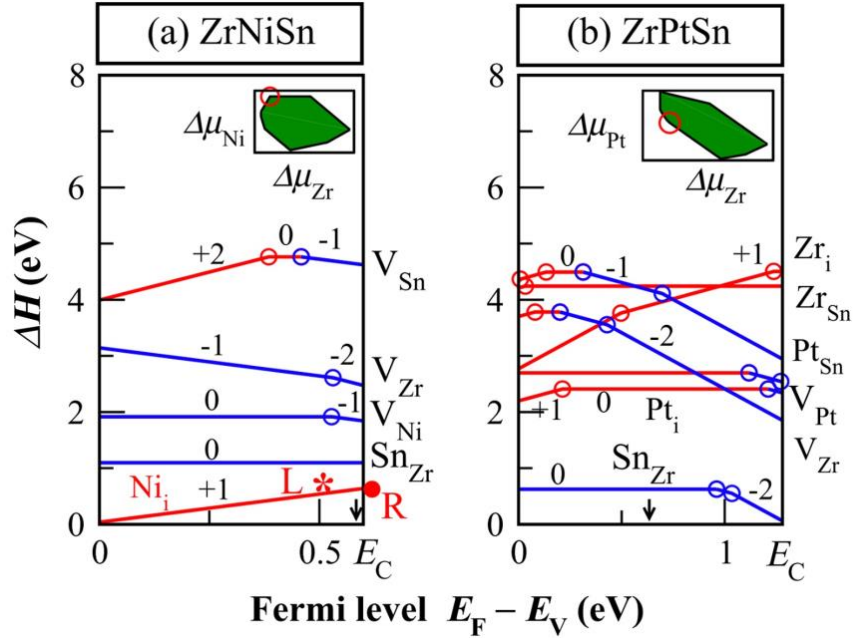

**Figure 9.** Defect formation energy (ΔH) as a function of Fermi level for (a) ZrNiSn and (b) ZrPtSn half-Heusler compounds. Positively charged electron-donor and negatively charged electron-acceptor defects are shown in red and blue, respectively. The Fermi level, valence band maximum, and conduction band minimum are denoted by $E_F$, $E_V$, and $E_C$, respectively. The inset demonstrates the green zone giving the range of chemical potentials where the compounds exist and specific chemical potentials used for calculations of ΔH versus $E_F$ within the allowed range of chemical potentials depicted by the green zone. The equilibrium Fermi level at growth condition (T = 850 °C) is shown by small vertical arrows. The transition levels are shown as open circles connecting two charge state lines. The figure is reprinted with permission from Ref. [72], Copyright (2017) by the American Physical Society.

Additional physical observations that can be gleaned from Eq. (1) regarding the dependence of doping energy on the nature of the host crystal is encoded in the total energy $E_H$ and band edge energy $E_V$. For example, (a) a host compound might be intrinsically nonstoichiometric, thus having a greater tendency to accommodate structural defects of the opposite polarity as its own. This can lead to spontaneous non-stoichiometry and self-doping, as described in Sec. 5.5. (b) Another example of the dependence of doping on the host compound relates to the possibility that some of its energy bands contain trapped carriers, i.e., polaronic-like bands (e.g., Fig. 3b,c). This can lead to anomalous "antidoping", as illustrated in Sec. 5.10. Finally, (c) the host crystal or its surface can be polar (as in LaAlO$_3$), contributing to the reduced formation energy of certain structural defects which then bias the intentional doping process, such as the efficient formation of electron-producing oxygen vacancy donors[189] at the surface of LaAlO$_3$, creating electrons that are swept to the SrTiO$_3$ region of the SrTiO$_3$/LaAlO$_3$ interface, explaining the interfacial conductivity[61], discussed in Sec. 5.11.

**4.4 Equilibrium Fermi level** is the thermodynamic Fermi level $E_F$ ($\mu_\alpha$, T) at temperature T and given chemical potentials $\mu_\alpha$ and depends self-consistently on all other defects and impurities. The equilibrium Fermi level $E_F^{eq}(T, \{\mu_\alpha\})$ is obtained from a coupled, self-consistent solution of the entire defect problem: all dopants and structural defects in all charge states are considered for their potential to contribute free carriers, along with thermal ionization of the host energy bands. The Fermi level $E_F$ is a continuous variable ranging between the VBM and CBM[72]:

$$\sum_{s=1}^{\text{nsite}} n_s \sum_{D,q} q_{s,(D,q)} \frac{1}{Z_s} \exp(-\beta \Delta H_{\text{form}}(D, q, \{\mu_\alpha\}, E_F)) = \int_{E_C}^{\infty} g(E) f_{\text{FD}}(E) dE - \int_{-\infty}^{E_V} g(E)(1 - f_{\text{FD}}(E)) dE, \quad (3)$$

where $\{\mu_\alpha\}$ are the chemical potentials; $n_s$ is the degeneracy of site *s* per unit cell; *g(E)* is the host crystal density of states obtained from the DFT calculation, and $Z_s$ is the partition function on site S due to all charge defects, i.e., $Tr(-\beta \Delta H_{\text{form}})$ with $\beta = 1/k_B T$.

The key point about the equilibrium Fermi level is its nonlinear dependence on *all charged species* in the material: (i) the formation energy $\Delta H(D, q, \mu_\alpha, E_F(i))$ is a function of the parametric Fermi level $E_F(i)$; (ii) the concentration $N_{D,q} (\mu_\alpha, T)$ of each defect depends on its formation energy, temperature, and chemical potentials; (iii) the *updated* parametric Fermi energy $E_F(i+1)$ is obtained from the charge neutrality condition, *summing over the contribution of all defects and dopants in all of their admissible charge states.* Consequently, this updated parametric Fermi level $E_F(i+1)$ now determines the updated formation energy of each defect and dopant. This shift between $E_F(i+1)$ and $E_F(i)$ will converge to a stabilized value and will thus determine the real equilibrium Fermi level $E_F^{\text{eq}}(T, \{\mu_\alpha\})$. But this dependence will also determine the pinning value of this Fermi level.

The formation energy of a specific defect depends on ALL defects*:* Because the energy to form a *given defect* depends on the Fermi level, and the latter is determined by ALL ionizable defects present. The calculation of the formation energy of a given defect (say vacancy) requires knowledge of all other charge-producing centers. This reflects the *"self-regulating"* character of the problem whereby chemical composition, defect charges, and local geometry of each defect are coupled in a feedback mode. Fig. 10 shows an example of the calculated equilibrium Fermi level for ZrNiSn and ZrPtSn as a function of chemical potential and growth temperature, indicating that ZrNiSn has its equilibrium Fermi level in the lower part of the gap (blue region in Fig. 10 left), whereas ZrPtSn has its Fermi level in the upper part of the gap (red region in Fig. 10 right).

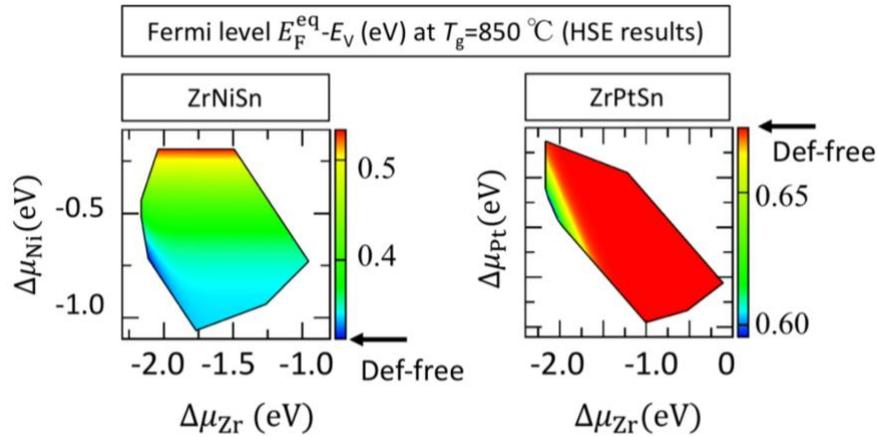

**Figure 10.** The equilibrium Fermi level with respect to the VBM at different chemical potentials (growth conditions) for n-type ZrNiSn and p-type ZrPtSn half-Heusler compounds due to intrinsic defect compensation. The figure is reprinted with permission from Ref. [72], Copyright (2017) by the American Physical Society.

**4.5. The equilibrium concentration of dopants and defects** $N_{D,q} (\mu_\alpha, T)$ and the free carrier concentration $N_e (\mu_\alpha, T)$ are functions of chemical potential and temperature. $E_F^{\text{eq}}(T, \{\mu_\alpha\})$ is calculated self-consistently. The net carrier concentration is the value from the right-hand side of Eq. (3). In experiments, samples are usually annealed at growth conditions for a prolonged period (usually days) before quenching to room temperature for measurements. In this case, one assumes that the amount of each individual defect type, which is the sum of all its charged states, was frozen in the lattice as in growth temperature, and only the ratio between individual charge states is allowed to vary at quench temperature obeying the Boltzmann distribution. Self-consistently solving the charge neutrality condition again leads to another set of solutions to $E_F^{\text{eq}}(T, \{\mu_\alpha\})$, and this time at a quenched temperature *T* from growth condition. The whole system re-equilibrates, and the resulting quenched carrier concentration at room temperature is shown in Fig. 11.

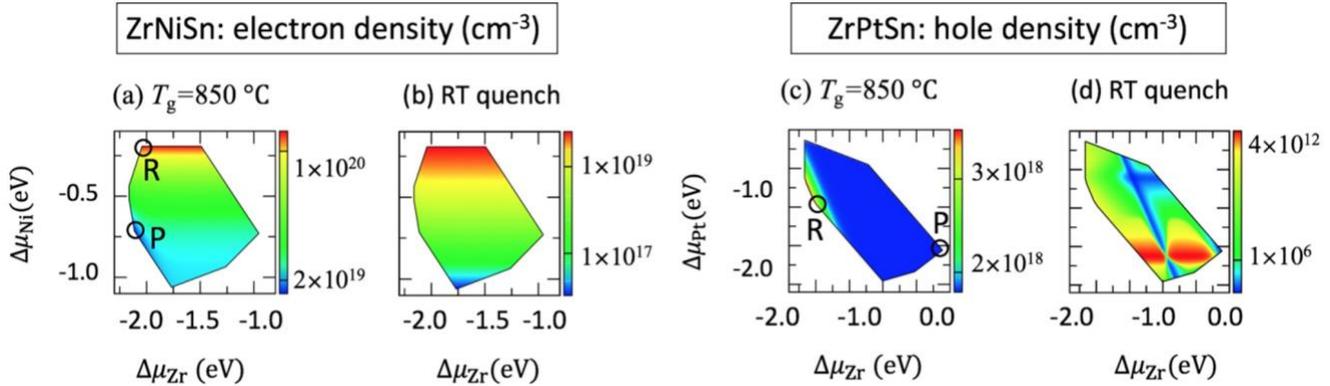

**Figure 11.** Majority carrier concentrations in two ABC half-Heusler compounds: *n*-type ZrNiSn and *p*-type ZrPtSn in the allowed chemical potential regime {$\Delta\mu_A$, $\Delta\mu_B$} at growth condition (high temperature) and at room temperature after quenched from growth. The letters R and P indicate rich and poor carrier concentration regime. The figure is reprinted with permission from Ref. [72], Copyright (2017) by the American Physical Society.

Using discussion in this section of the six key quantities implicated in doping physics as illustrated in Fig. 7-11, we next explain the doping trends and peculiarities noted in Sec. 2 with the explanation of the underlying concepts of the doping of Sec. 4.

## 5. THEORETICAL DOPING CONCEPTS EXPLAIN DOPING PHENOMENOLOGY AND DOPING PECULIARITIES

**5.1 The self-regulating response: Doping creates antibodies to doping:** One may naïvely think that the introducing of free carriers to the system moves the Fermi level to the valence band for hole doping while the electron doping shifts the Fermi level to the conduction band, and that the introducing of more free carriers simply requires only addition of more dopants. However, as explained in Sec. 5.4, the position of Fermi level is defined by the charge neutrality rule (Eq. (3)) and should be calculated by accounting for all possible defects. Indeed, intrinsic *structural defects* might affect the intended doping. For example, the removal of anions from a lattice (forming anion vacancies) forms neighboring cation dangling bonds with *less* than a half-filled shell and thus tends to be intrinsic *electron donors*, whereas the removal of cations (forming cation vacancies) forms anion dangling bonds having *more* than a half-filled shell, thus tends to be intrinsic *hole-producing* acceptors. Since the formation energy of charged defect depends linearly on the Fermi level, doping can result in a self-regulating response – the spontaneous formation of intrinsic defects that tend to negate the polarity of the deliberately created dopant. This effectiveness of the self-regulating response reflects the propensity of the pristine compound to form structural defects of a given charge. Consider, for example, a spinel $A^{3+}_2B^{2+}O_4$ such as $Co_2ZnO_4$, where the B-on-A anti-site defect is an acceptor, whereas the A-on-B anti-site is a donor.[89-92] The order and energy separation between the corresponding acceptor and donor charge transition levels can vary. Figure 12a shows different possible energy ordering of anti-site acceptors and anti-site donors. For example, type 1 (T1) spinels has anti-site acceptor- below- donor transition levels and both located within the band gap, whereas types 2, 3, and 4 spinels show the order of donor- below- acceptor with different energy shifts, e.g., type-4 spinel has both levels in the band gap. Figure 12b shows DFT calculations[89] of defect levels for $Co_2ZnO_4$ indicating that the Co-Zn pair forms a T2 case.

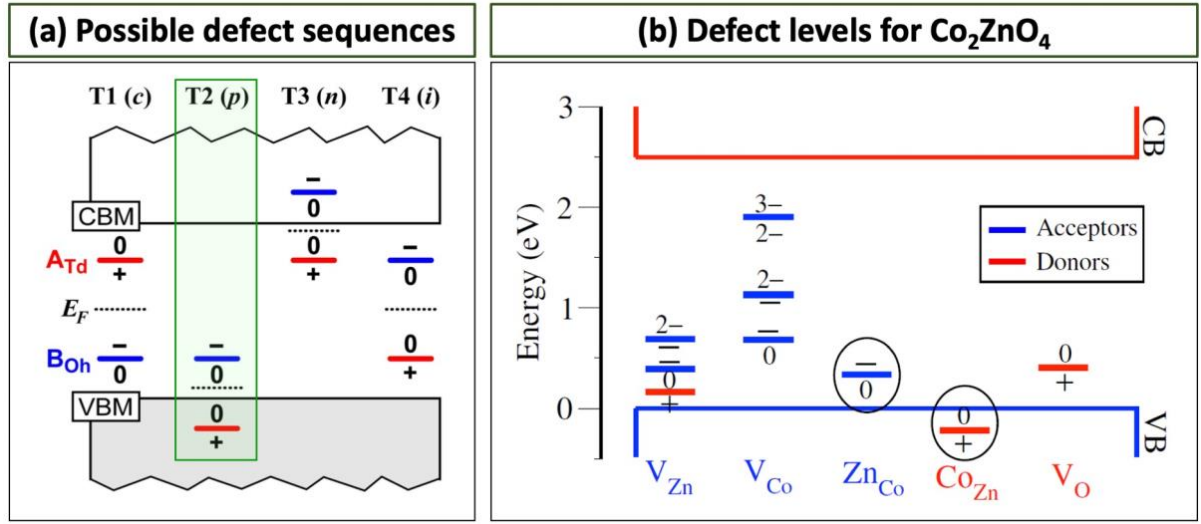

**Figure 12.** (a) Schematic illustration of the order of transition levels for the donor A-on-Td (red) and the acceptor B-on-Oh (blue) where $T_d$ and $O_h$ are the tetrahedral and octahedral sites. The order on such transition levels defines four types (T) with doping characteristics: "compensated" (c), p-type (p), n-type (n), or intrinsic (i). (b) Actual DFT calculations for defect levels in $Co_2ZnO_4$, showing that this compound belongs to type T2, i.e., Co-on-Zn donor in the valence band energetically lower than the Zn-on-Co acceptor. The figure is redrawn using data from Ref. [89].

Let us illustrate how the order of transition levels of opposite polarity in a given compound can decide if the Fermi level can be tuned by doping. Assuming for simplicity that the formation energies of the acceptor and donor metal-on-metal anti-sites are roughly comparable, it is possible to distinguish the different behaviors of the compound where $E_F$ can be tuned over a wide range, vs. those where the tuning of $E_F$ is limited to a narrow range. Figure 13a,b shows for the case of ***donor- above- acceptor*** the formation energy of defects vs. the parametric Fermi level, ranging between the VBM and CBM**,** leading to a narrow region (yellow highlight) over which $E_F$ can be tuned with doping. Now, if one dopes deliberately by electrons, moving in Fig. 13b towards the left-hand side (n-type doping), the Fermi level will rise due to electron addition. However, then the formation energy of the acceptor anti-site will be lowered (blue line in Fig. 13a), creating more holes that in turn compensate the deliberately added electrons. Such response slows down the increase of the Fermi level, causing pinning of the Fermi level for n-type doping (Fig. 13b). This condition represents the Fermi level, where the formation energy of the acceptor "killer defect" vanishes. Conversely, doping deliberately by holes (Fig. 13b right-hand side, p-type doping) would lower $E_F$ towards the VBM until the formation energy of the donor killer defect (that produces electrons, compensating the holes) vanishes (red line in Fig. 13a). This defines the pinning Fermi level for p-type doping. Because the formation energy of donors and acceptors cross in this case in mid-gap (Fig. 13a), the range of Fermi level tuning is limited by the feedback effect. In the reverse level order of **donor-below acceptor** (i.e., donor closer to the VBM, T4 type in Fig. 12a), where the formation energy of donor and acceptor do not meet in mid-gap (Fig. 13c), the Fermi level can move quite a bit (yellow highlight in Fig. 13 c,d) up by deliberate n-type doping before it encounters the acceptor anti-site that will pin it down.

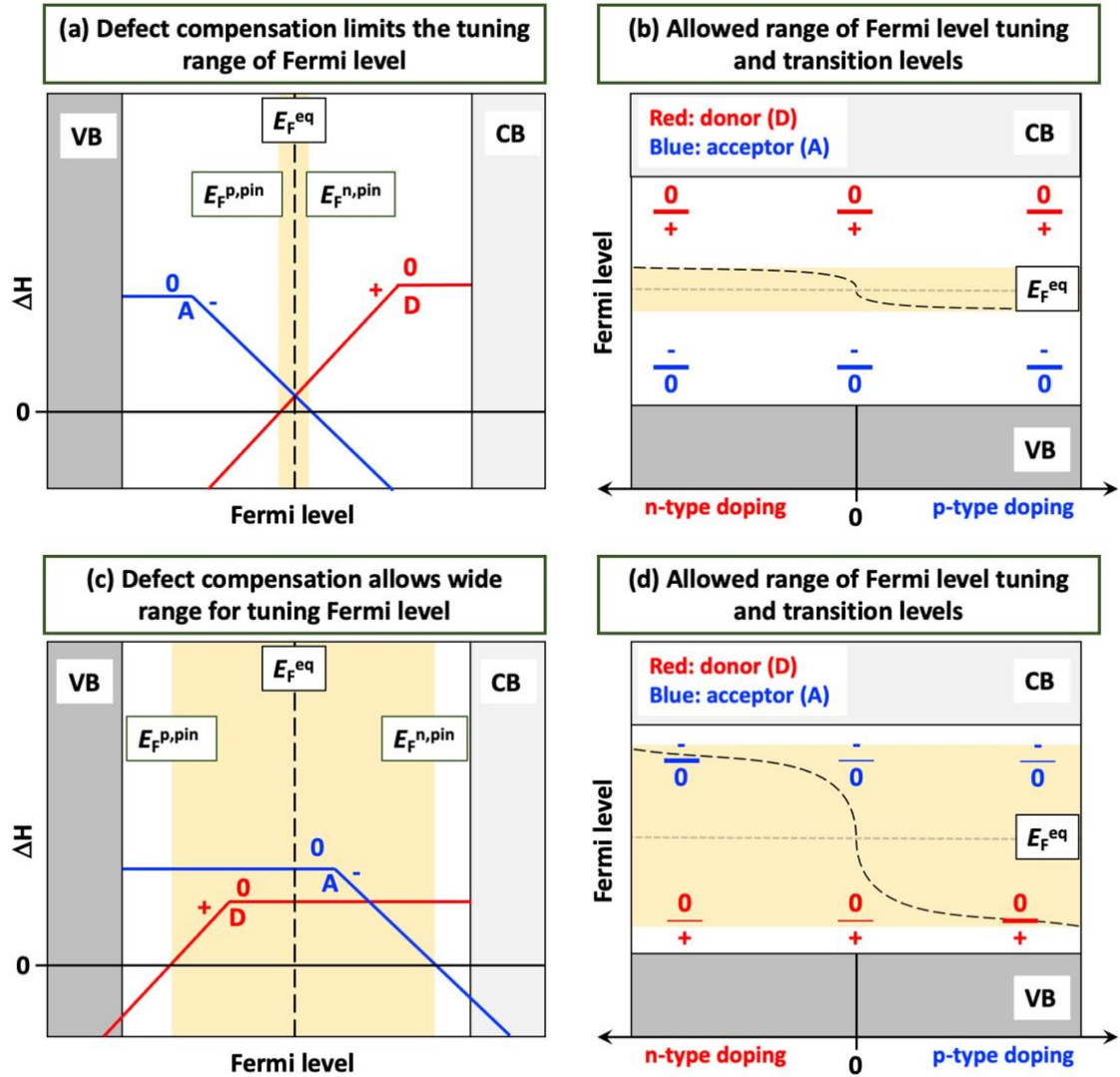

**Figure 13**. Schematic illustration for (a,b) donor- above- acceptor (donor level closer to the CBM) and (c,d) donor-below acceptor transition levels. Possible range for tunning Fermi level shown in yellow is defined by upper ($E_F^{n,pin}$) and lower ($E_F^{p,pin}$) Fermi level pinning. The equilibrium Fermi level is denoted as $E_F^{eq}$. The figure is redrawn from Ref. [89].

Figure 14 shows another illustration of self-regulation in doping, using the DFT calculations on n-type doping of ZnO by Al and the case of Sn doping of $In_2O_3$. Adding n-type dopants to ZnO shifts the Fermi level towards the CBM, increasing the free electron density in direct proportion to the incorporated Al.[63] However, at some critical concentration of free electrons, the increase of $E_F$ with Al doping starts leveling off. Inspection of the concentration of Zn vacancies in the supercell (lower panel of Fig. 14a) clarifies that the pinning of the Fermi level is associated with a marked increase of $V_{Zn}$ concentration, an acceptor that compensates electrons, poisoning the doping reaction. The same observation is predicted for Sn doping of $In_2O_3$, as shown in Fig. 14b. The response of ZnO and $In_2O_3$ to electron doping is summarized in Fig. 14c.

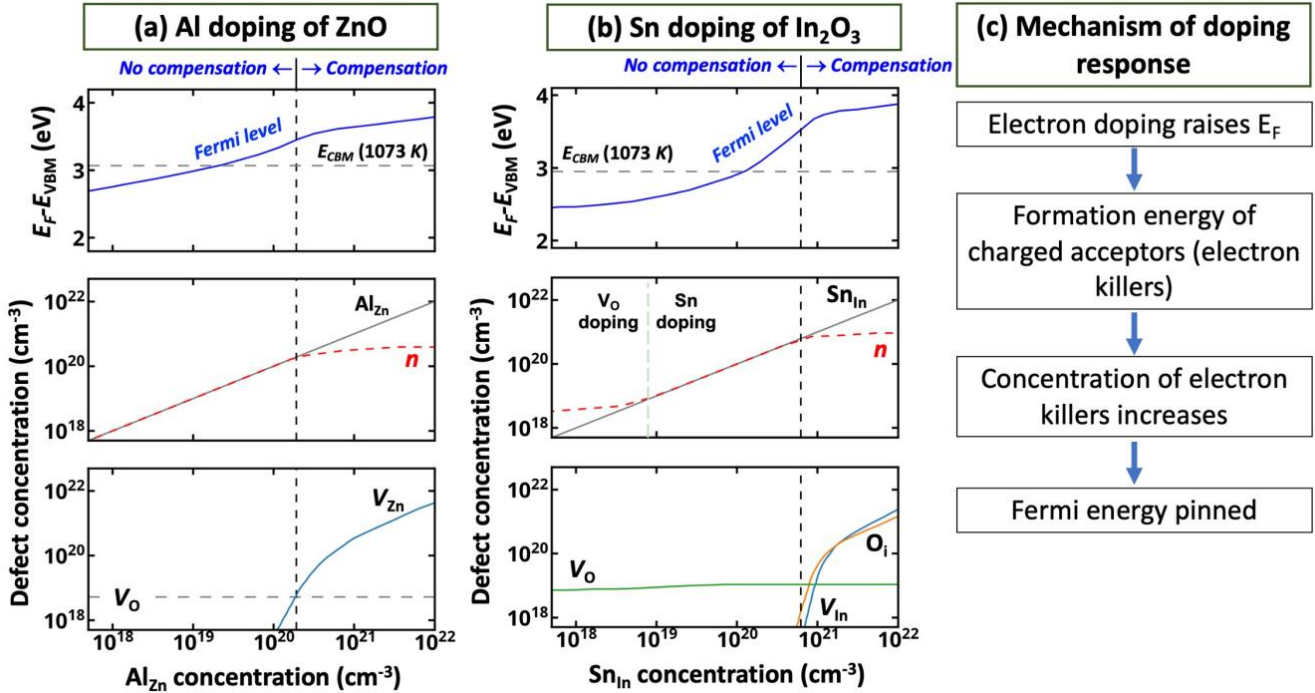

**Figure 14.** Defect compensation limits n-type doping of (a,c) ZnO by Al and (b,c) $In_2O_3$ by Sn. At low doping concentration, n-type doping results in the introduction of free carriers and the concomitant increase in $E_F$. Starting from a certain n-type doping concentration, charged acceptors (e.g., $V_{Zn}^q$ in ZnO and $O_i^q$ in $In_2O_3$) are formed in response to doping, acting as killer defects limiting further electron doping. The figure is redrawn using data available in Refs. [63,190].

**5.2 Asymmetry between tendencies for electron vs. hole doping reflects different efficiencies of the self-regulating response:** Figure 15 shows band alignment with n- and p-type pinning energies for simple II-VI oxides[79] whereas Fig. 16 depicts earlier results[24,26,84] for a series of III-V and II-VI compounds. For ZnO, the n-type pinning energy in Fig. 15 is inside the conduction band, so plenty of electrons can be doped in before $E_F$ reaches this pinning energy, at which point further n-type doping will come to a halt. The ZnO p-type pinning energy is generally above the VBM (this depends, however, on the growth condition favoring or disfavoring the formation of the hole killers, i.e., anion vacancies, Fig. 16) suggesting that it is difficult to realize ZnO with the Fermi level in the valence band because the killer defect will form efficiently. In NiO, on the other hand, the p-type pinning energy is inside the valence band, so plenty of holes can be introduced by doping before their demise by oxygen vacancies. But the n-type pinning level of NiO is well below the CBM, implying that a large concentration of electrons cannot be reached by doping. This behavior is similar to that for MgO, where the high-concentration of hole doping can be introduced owing to p-type pinning energy significantly lower than the VBM, however, electron doping is limited by $E_F^{n,pin}$ almost 3.3 eV below the CBM. These tendencies in the limits of dopability reflect a fundamental asymmetry in the bonding of the host compound, some being more favorable to created electron producing structural changes, some more amenable to create hole producing structural changes. The rather deep VBM in ZnO relative to the higher VBM of NiO reflects the greater level repulsion between Ni-d and O-p than the weaker repulsion between the Zn-d and O-p. Figure 16 shows a similar prediction by DFT for III-V and II-VI compounds[24,26,84] where the individual calculated pinning energies were averaged by the respective horizontal lines. Considering the upper n-type pinning energy, we see that most III-V and ZnO, CdS, CdSe, and CdTe have the n-type pinning energy inside the conduction bands, so they are n-type dopable. Considering the p-type pinning energies, we see that cases with this energy inside the valence bands include III-V antimonides and II-VI tellurides being readily p-type dopable, whereas cases, where the p-type pinning energy does not penetrate the valence band, include III-V Phosphides and II-VI oxides and sulphides. We note that these doping tendencies reflect

equilibrium growth and that manipulating growth to suppress kinetically the formation of structural killer defects such as oxygen vacancies or metal vacancies) would extend dopability beyond the equilibrium limits.

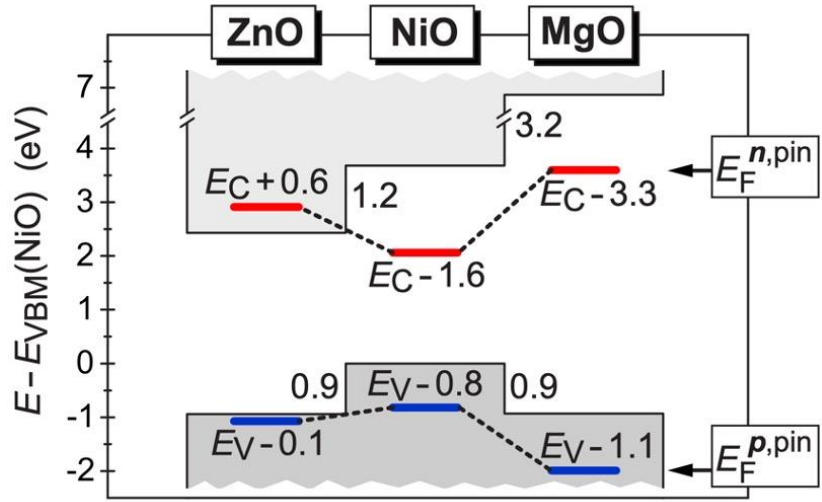

**Figure 15** Band offsets between ZnO, NiO, and MgO with the respective upper ($E_F^{n,pin}$) and lower ($E_F^{p,pin}$) limits of Fermi-level pinning energies defined by compensating defects. $E_F^{n,pin}$ and $E_F^{p,pin}$ are determined for O-rich and metal-rich conditions. The figure is reprinted with permission from Ref. [79], Copyright (2007) by the American Physical Society.

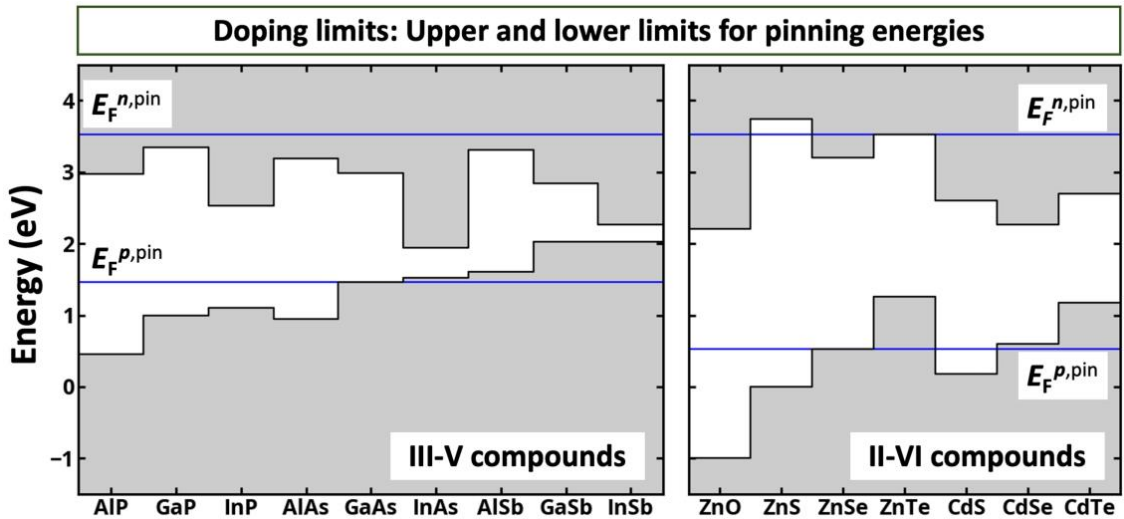

**Figure 16** Band offsets between III-V and II-VI compounds with the respective upper ($E_F^{n,pin}$) and lower ($E_F^{p,pin}$) limits of Fermi-level pinning energies defined by compensating defects. The figure is redrawn using data from Ref.[84].

**5.3 Doping to the target Fermi level:** One often reads in theory articles on Quantum Materials that the authors express the wish that the Fermi level be shifted to a desired location in the band structure to achieve a particular electronic effect (e.g., topological band inversion[33]). For example, Yan et al.[33] suggested converting an ordinary non-topological compound $BaBiO_3$ into a topological insulator by shifting the Fermi level considerably upwards. In the follow-up exploration, Khamari and Nanda[191] suggested the potential experimental verification for such exotic properties by F doping and realization of $BaBiO_2F$ compound in the cubic space group. Many insulators are said to become high $T_c$ superconductors if only $E_F$ can be placed inside the valence or conduction bands. "Doping to the Fermi level" can be used to stabilize different charge state of point defects for specific applications (e.g., silicon-vacancy (Si-V$^q$) color center in diamond[192,193]) or tune the critical properties such as superconductivity of the metallic materials to the optimum region[185,194]. Also, doping of *metallic compounds* has been studied experimentally for high $T_c$ superconductors[195,196], field-effect transistors[197], and plasmonics.[198] Furthermore, ideas about devices based on extracting spin-polarized electrons from Rashba spin-split bands in an insulator requires that these Rashba bands be located near the VBM or CBM; otherwise, doping to a target position would be needed.

Our foregoing discussion indicates that in order to dope to a target Fermi level, one needs to be mindful of three bottlenecks:

*First,* as indicated in Figs. 15 and 16, each compound has an intrinsic Fermi level pinning energy, whereby attempts to dope past that level start producing intrinsic defects that halt further doping. Often, an intentional *n*-type pinning level is associated with spontaneous production of acceptors (*electron killers)* such as cation vacancies, and low-valent anti-site defects, whereas *p*-type pinning level is associated with spontaneous production of *hole killers* such as anion vacancies and some high-valent anti-site defects*.*

*Second,* by shifting the Fermi level, doping can destabilize the undoped crystal structure and stabilize another structure type. For example, the 2*H* polytype of transition metal dichalcogenides can change to the 1*T* polytype once the Fermi level is raised[199,200] or Wurtzite vs. Zinc blende phase transition in GaN due to doping.[201] The transformed structure may have different doping tendencies.

*Third,* elevating the Fermi level can lead to the occupation of predominantly anti-bonding levels, thus destabilizing the structure in favor of another one which, alas, may not have the target topological band inversion. For instance, the cubic phase of $BaBiO_3$ is not the ground state and under normal conditions reconstruct to monoclinic phase. When the lower energy monoclinic system is doped, the resulting system becomes a trivial insulator without band inversion[82,83]. Similar behavior is observed in $BaBiO_2F$, where the predicted cubic phase is highly unstable (by 285 meV/atom) with respect to the decomposition to $Bi_2O_3$, $Ba_2Bi_2O_5$, and $BaF_2$.[82]

These examples indicate that a valid prediction of whether a target position of Fermi level is realizable could benefit from a total energy calculation of the system with shifted Fermi level, examining its structural stability.

**5.4 "Doping rules" and "designer dopants":** Another way of looking at the pinning energies in Fig. 15 and 16 leads to the formulation of Pauling-esque "doping rules"[84] that can help us navigate rationally throughout the Periodic Table of doping. For example, we see from Fig. 16 that compounds with low energy CBM (high bulk electron affinity) can be generally doped by electrons before electron killing structural defects set in, whereas compounds with high energy VBM (low bulk work function) can be readily doped by holes, before hole killing structural defects form. That n-type doping is enabled more readily in compounds with low energy CBM (large electron affinity), such as many oxides and sulphides where low lying cation s orbitals form the CBM, whereas p-type doping is enabled more commonly in compounds with high energy VBM (i.e., compounds with large Z anions such as tellurides or antimonides) is, in fact, known in ordinary insulators, but not sufficiently explored in more contemporary Quantum Materials. A word of caution on generalizing such rules: We note that the VBM is not necessarily anion p-like, which is common for traditional oxide insulators. For example, in rocksalt CoO, FeO, and $ABO_3$ perovskites with early transitional 3d metals such as $YTiO_3$ and $LaVO_3$, the VBM is orbitally metal d-like, so the nature of the structural defects that halt p-type doping may be different.

This type of understanding of doping bottlenecks can lead to the selection of compounds that can circumvent doping bottlenecks. Oxides are often difficult to dope because doping can create spontaneously structural defects that negate the intentional doping. For example, in the $A^{3+}_2B^{2+}O_4$ spinels, the substitution of $B^{2+}$-on-$A^{3+}$ (low-valence cations on a high valence cation site) forms a *hole-producing* acceptor level $\varepsilon(0/-)$. However, the formation of the opposite $A^{3+}$-on-$B^{2+}$ anti-site creates a *hole-killer* donor $\varepsilon(+/0)$. The latter defect would normally halt intentional p-type doping. But if one can identify a compound where the said anti-site donor level is not located in the band gap, but rather in the valence band, the said compound could be doped p-type without fear of compensation by the donor anti-site. An exciting strategy will be to search for compounds where the opposing ("killer") defect transition level lies outside the band gap. For example, Paudel et al.[89] established that in $Co_2ZnO_4$ (Fig. 12b) the donor transition level $\varepsilon(+/0)$ lies below the VBM, so in this case intentional p-type doping does not encounter opposition. This was indeed verified by Perkins et al.[90] showing that the p-type doping could be increased significantly by (i) increasing of $Zn^{2+}$ concentration (via nonequilibrium growth) thus raising acceptor anti-site concentration and by (ii) inducing the "inverse spinel" phase to maximize divalent cations on octahedral site, without fear that donor anti-sites will compensate the acceptors because the former are electrically inactive. Such examples illustrate the option of "designer doping" based on understanding of the limiting mechanisms.

**5.5 Doping by non-stoichiometry:** The half-Heusler compounds are known for their natural off-stoichiometry and, apparently independently, are also known for being naturally doped. We have used the compounds ZrNiSn and ZrPtSn as prototype systems to illustrate the doping principles in Sec. 4. The calculations of chemical potentials (Fig. 8) and defect formation energies (Fig. 9) showed the abundant low formation energy defects (e.g., Ni interstitial being an electron producing donor in ZrNiSn), resulting in free carrier concentration (Fig. 11) being $\sim 10^{19}$ cm$^{-3}$ and $\sim 10^{18}$ cm$^{-3}$ in ZrNiSn and ZrPtSn, respectively. Indeed, the compounds containing 3d elements in the B position (ZrNiSn) is naturally B-rich and n-type[73], whereas the compounds containing 5d elements in the B position (ZrPtSn) is often C and A element-rich and p-type[95].

Understanding this type of connection between natural, non-Daltonian off-stoichiometry and doping tendencies might be particularly interesting in classic topological insulators, such as $Bi_2Se_3$[202] that tend to be off-stoichiometric (Se-deficient) and are also known to have their Fermi level located within the conduction band. This makes them practically bulk metals instead of the intended (topological) insulators where the Dirac cones are ideally supposed to be in the band gap. However, defect calculations[203] showed that this specific natural off-stoichiometry is the *cause* of being also electron-rich, and suggested artificial, designer doping to counter this tendency.

**5.6 Symmetry lowering by spontaneous defect formation can destroy exotic topological features:** Spontaneous formation of vacancies could alter the electronic structure, doping the material and thus displacing the Fermi level to a region that no longer has the exciting properties of the original structure.

The generic electronic structure illustrated in Fig. 1a corresponds to a metal with Fermi level in main conduction band; however, unlike more innocent-looking metals, this electronic configuration has a significant "internal band gap" below the CBM, opening the possibility for an energy-lowering electronic instability that would create a split-off band inside the internal gap (Fig. 1b). We have seen that this is the case for LaMnO$_3$ (Fig. 5b). The p-type analogue, where the initial state has E$_F$ inside the principal *valence band*, is tetrabarium tribismuth Ba$_4$Bi$_3$ in space group 220. The fact that, in this configuration, E$_F$ is inside the principle valence band[29] indicates that this material is missing electrons. The band structure in the metallic state has a special 3-fold and 8-fold fermions below the Fermi level[29], a consequence of specific non-symmorphic symmetry existing in the space group 220. This unusual band symmetry is, in fact, a recently predicted unconventional quasiparticle[29] that remarkably has no analogy in particle physics, having created recently significant interest. However, DFT calculations[82] showed that the creation of Bi vacancy is rewarded chemically by lowering the energy of this crystal — it releases electrons that fill the missing electron slots in the valence band. Alas, this predicted exothermic process removes the original lattice symmetry that was predicted[29] to create the exotic quasiparticle. While it is entirely possible that the new lattice structure would have another (perhaps mundane) topological symmetry, the original exotic quasiparticle symmetry is gone.

**5.7 The coexistence of transparency and conductivity as a special form of doping with impunity:** In the physics of quantum materials, there are rare compounds that exhibit coexistence of often mutually exclusive properties of high conductivity (usually common only in opaque metals) and optical transparency common (usually in wide-gap insulators). Historically, such compounds were realized by heavy doping of wide band gap insulators, e.g., p-doping of CuAlO$_2$[64] and Cr$_2$MnO$_4$[65] or n-doping of In$_2$O$_3$[66] and ZnO[63]. However, as explained above, not all insulators can be successfully doped. Only if the n-type pinning energy is above the CBM, the systems can be successfully doped n-type and be a potential transparent conductor. If the p-type pinning energy is in the valence band, the system can be successfully doped p-type.

Recently, Zhang et al.[156] demonstrated an alternative approach to design the novel transparent conductor by starting from the intrinsic degenerate gapped metals (e.g., schematically shown in Fig. 1a) and trying to make it transparent. This approach has been later experimentally demonstrated for $SrVO_3$ and $CaVO_3$.[136] The advantage of such methods lies in the possibility to have ultra-high free carrier concentrations (e.g., over $10^{19}$ cm$^{-3}$) as well as avoid the complex process of doping. Moreover, owing to free carriers in the conduction or valence band, the properties of such unique systems can be tailored via controllable non-stoichiometry, meaning that specific transparency, conductivity, and phase stability can be selected by zooming to specific chemical potentials (growth conditions). The point is that compounds that have Fermi level in the principal conduction band can have the spontaneous formation of acceptor vacancies due to decay of conducting electrons to the acceptor level (compare Fig. 1a with Fig. 1b). Eventually, this ability can lead to violating Daltonian stoichiometry and formation of non-stoichiometric compounds having a large concentration of acceptor vacancies, which in the concentrated limit can condense into ordered crystallographic arrays[157]. For instance, in $Ba_lNb_mO_n$ family of compounds, the set of 1:2:6, 3:5:15, 5:4:15, 7:6:21, 7:8:24, 9:10:30, and 26:27:81 ordered vacancy compounds has been found.[157] Moreover, each of such compounds is stable with respect to decomposition to competing phases and has a specific value of free carriers and own optical properties. This discovery thus demonstrated how non-stoichiometry can be an intrinsic effect of a compound and not a growth artifact and might have a direct effect on symmetry-enabled quantum properties.

**5.8 Electronic impurities: Polarons and split-off in-gap intermediate bands:** The traditional view of doping was atomistic and based on the notion that an impurity **I** substituting a host atom **H** creates a perturbation $\Delta V_{imp}$=V(I)-V(H). When the magnitude of such perturbation exceeds a threshold value, an impurity level would form within the band gap region. This atomistic picture, anchored in the pioneering Greens function model of Koster and Slater[96] and its extension to 3D semiconductors by Hjalmarson et al.[97] directed the field of impurity doping to consider atomistic constructs such as I vs. H size mismatch, electronegativity mismatch, etc. It was noted, however, that such doping levels could form inside the band gap, just as a pure electronic event without a substitutional impurity atom that is integrated into the lattice. This can be achieved by introducing carries without the impurity being strongly interacting or even present, such as in photodoping, modulation doping, gating, or by creating a weakly coupled *interstitial* impurity (Li, H) that gets ionized without being integrated into the lattice structure. The two leading categories of electronic impurities include (a) polarons (e.g., electron polaron in e-doped $TiO_2$[98-100] or hole polaron in h-doped $SrTiO_3$ [6,201]) whereby a localized state is formed in the gap such as $Ti^{3+}$ state with wavefunction localized on *single* reduced cation Ti site, and (b) formation of split-off in-gap electron or hole intermediate *band*s such as in $YTiO_3$ and $LaTiO_3$. These two classes will be discussed next.

*(a) Electronic perturbations leading to the formation of polaron state*: This is illustrated by interstitial Li in monoclinic $TiO_2$ (SG: 12) where each Li is donating an electron to the Fermi sea. This electron is trapped by Ti converting one $Ti^{4+}(d^0)$ into $Ti^{3+}(d^1)$, thus resulting in the formation of an occupied in-gap state (Fig. 17a,b). A similar example of "hole impurity" can be illustrated by the hole polaron in $SrTiO_3$.[6,201]

One of the main failure modes of low mobility in oxides is the spontaneous formation of polarons associated with local lattice deformation.[204] Yet, for a long time, it has been impossible to forecast by DFT which oxides have the propensity to self-trap polarons and which do not. The reason being is the built-in tendency in the standard DFT XC functionals to lower the total energy as wavefunctions spread out, a tendency apparent in the *downward* bowing of the total energy with occupation number. New developments in DFT now permit systematic calculation of polaron binding energy[205], thus affording rational selection of suitable oxides that do not impede mobility by forming (small) polarons. The method called the "Cancellation Of Non-Linearity" (CONL) eliminates or at least reduces the self-interaction error in conventional DFT and recovers the linear behavior of electron total energy versus occupation number, fulfilling the Koopmans' theorem. The method is extremely efficient, 1000 times faster than the hybrid functionals, but has the power to correctly predict unbound hole polarons in $Rh_2ZnO_4$ [206] and self-trapped polarons associated with O-p in $TiO_2$ [207], and was recently used to screen polaron binding energy in ternary

Mn(II) oxides[65]. The CONL correction is done by adding an onsite electron potential on certain *lm*-decomposed orbitals on top of the DFT+U framework. The electron state potential for electron doping is

$$V_e = \lambda_e (p_{m,\sigma}/p_{host} - 1), \tag{4}$$

where $p_{m,\sigma}$ and $p_{host}$ denote the target hole occupation of the $m$ sublevel of spin $\sigma$ and the target hole occupation of the host material without doping. The constant $\lambda_e$ is determined by requiring the Koopmans condition, i.e., that the difference in total energies for N and N-1 electrons will equal the respective eigenvalue, thus obeying the generalized Koopmans condition. Without this term, ordinary DFT tends to lower its total energy when orbitals *spread out* (causing universal instability of polarons), just the reverse of Hartree-Fock (HF) theory that lowers the energy when orbitals become *more compact* (causing universal stabilization of polarons). This is the reason that a judicious mix of DFT and HF (as realized by the hybrid functional approach) generally gives reliable polarons.

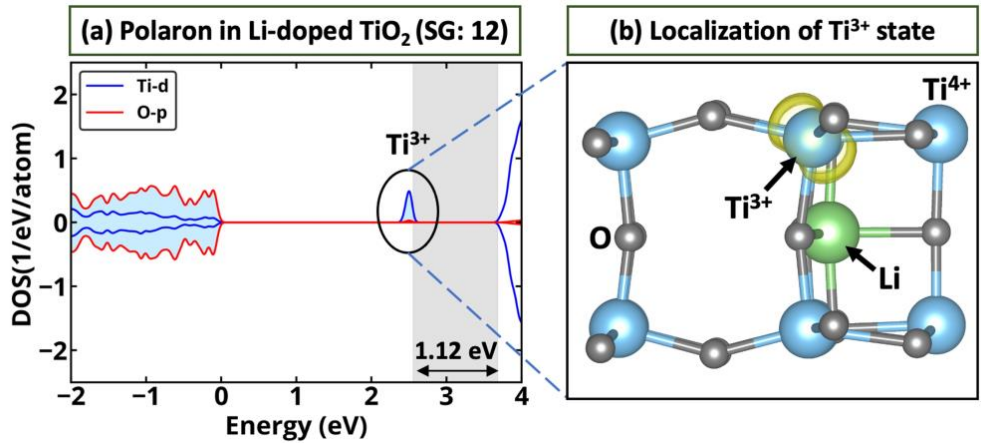

**Figure 17.** Formation of localized polaron states in monoclinic $TiO_2$ (SG: 12) as a result of doping. (a) Doping of monoclinic $TiO_2$ (SG: 12) by Li resulting in the formation insulator with an electron-trapped state. (b) The wavefunction squared corresponding to the localized in-gap state showing that Li doping resulted in the formation of a reduced $Ti^{3+}$ state localized on a single Ti atom. Occupied states are shadowed in light blue. The band gap of the system (shown in silver) ranges from the valence band to the lowest energy unoccupied level. The figure is redrawn using data from Ref. [99].

***(b) Electronic perturbation leading to the formation of a split-off in-gap intermediate band with trapped carrier:***
A related case of electronic impurity that involves the formation of a full band (not just a small cluster of ions as in a polaron) called a split-off in-gap band is illustrated in Fig. 18a. Whereas the *nonmagnetic* $LaTiO_3$ and $YTiO_3$ described by naïve DFT with all 3d ions being equivalent, thus predicting a (false) metallic state, the real paramagnetic state (where each 3d ion can have a different local moment, all adding up to zero[41-43]) results in the formation of in-gap states. Indeed, the insulating nature of both compounds with distinct in-gap states is known experimentally.[208-211] The analysis of wavefunction squared corresponding to the in-gap states demonstrates that these states are localized on metal sublattice and hence can be discussed as $Ti^{3+}$ (Fig. 18b). The perturbation inducing such localization can be illustrated by the difference of Hartree potentials for supercells having in-gap state and not having in-gap state, indicated symbolically as $\Delta V_{elec}$= V($Ti^{3+}$) - V($Ti^{4+}$), i.e., same atoms with different electron distribution. Indeed, as Fig. 18c shows, this electronic perturbation – the difference between the Hartree potentials of the paramagnetic and nonmagnetic phases computed for the same (unrelaxed) supercell— demonstrates a clear driving force for electronic symmetry breaking, in remarkable analogy with atomic impurities where the difference between host and impurity potentials $\Delta V_{imp}$=V(I)-V(H) creates a split-off state in the gap. In addition to the split-off state that contains a trapped electron, there are split states trapping holes in undoped $YNiO_3$ and $SrBiO_3$.[53]

Interestingly, unlike insulating LaTiO$_3$, YTiO$_3$, and YNiO$_3$, the compound SrVO$_3$ being a degenerate gapped metal with Fermi level in conduction band stays as a metal even in paramagnetic calculations[163]. The reason is that it is stable in the cubic phase (as hinted by its close to unity Goldschmidt factor) and as such does not have octahedral rotations that could otherwise open the gap. These results thus further imply that structural symmetry breaking is an essential part of band gap opening in ABO$_3$ perovskites as has also demonstrated by Varignon et al. [42,43]

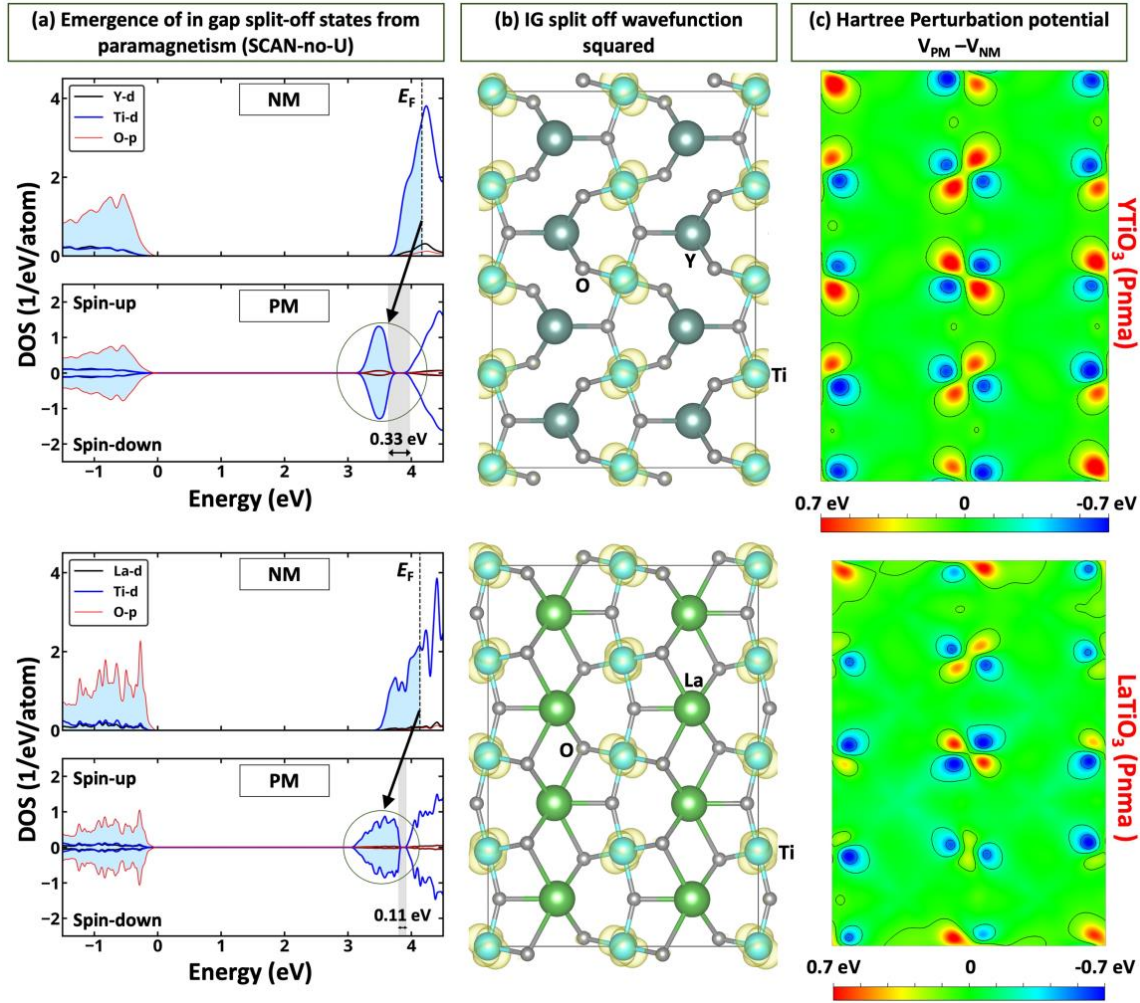

**Figure 18.** Formation of split-off in gap intermediate band trapping electrons in the paramagnetic YTiO$_3$ and LaTiO$_3$ using SCAN-no-U. (a) Density of states for Pnma paramagnetic (PM) YTiO$_3$ and LaTiO$_3$ showing the appearance of in-gap states when local magnetic moment is allowed as compared with nonmagnetic (NM) approximation. Occupied states are shadowed in light blue. (b) Wavefunction squared corresponding to the in-gap (IG) states. (c) Perturbation of electrostatic potential computed as the difference of Hartree potentials for paramagnetic and nonmagnetic systems having the same internal structures plotted for (010) plane of the 160-atom supercell. The band gap is shown in gray. SG denotes the space group number.

*(c) The analogy between split-off bands in symmetry-broken DFT and split-off states in correlated models of Mott insulators:* We have seen in Figs. 3, 4, 5, and 18 the illustration of how metallic electronic structure is transformed under the effect of some "electronic perturbation" such as V$_{elec}$ into a reconstructed electronic structure where the broad initial metallic states create a split-off state. In Figs. 3 and 18, allowing for metal atoms to have local magnetic moment results in band gap opening and formation of empty *split-off hole-trapping band* for YNiO$_3$ and occupied *spit-off electron-trapping bands* in YTiO$_3$ and LaTiO$_3$. Similarly, energy lowering

disproportionation and pseudo-Jahn Teller $Q_2^+$ distortion causes the band gap opening and formation of electron-trapped split-off states in SrBiO$_3$ (Fig. 4) and LaMnO$_3$ (Fig. 5), respectively. These split-off states appearing in mean-field (but symmetry broken) DFT are reminiscence of the split-off states described by Sawatzky et al.[22] in the language of strongly correlated physics (Fig. 19). Doping of normal semiconductors (compounds, where the valence band is defined usually by ligand p states and conduction band minimum is metal s-like) will result in a simple shift of Fermi level, e.g., adding electrons will place Fermi level inside the conduction band, while hole doping will result in Fermi level being inside the valence band but no split–off states. This semiconductor behavior is the commonly seen in previous DFT calculations of doping semiconductors.[63,190] Fig. 19 Illustrates further that bands composed of metal d electrons—such as Mott-Hubbard insulators or charge transfer insulators (e.g., compounds with VBM defined by ligand p states and the conduction band edge by metal d states) will split state of the conduction band into the gap in response to electron doping. This is analogous to DFT results illustrating electron split-off bands in Fig. 18 for LaTiO$_3$ and in YTiO$_3$ or for Li-doped TiO$_2$ in Fig. 17. It is evident that symmetry broken, mean-field DFT describes the same phenomenology of split off states attributed previously exclusively to dynamic correlation.

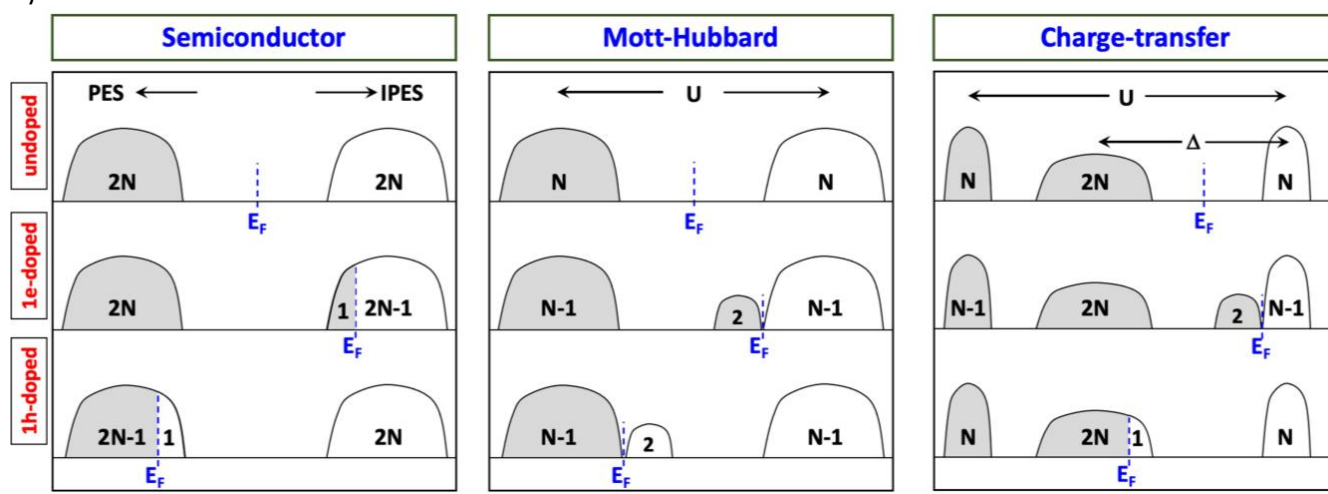

**Figure 19.** Model Hamiltonian expectation of the consequences of doping semiconductor (e.g., VBM and CBM are defined by p states and s states, respectively), or doping Mott insulator (e.g., both VBM and CBM are defined by d states), or doping charge transfer compounds (e.g., VBM and CBM are defined by p states and d states, respectively). Split-off bands are indicated. The on-site repulsion, charge-transfer energy, photoemission, and inverse photoemission are labeled as U, Δ, PES, and IPES, respectively. The figure is redrawn from Ref.[22].

**5.9 When isolated localized insulating polarons coalesce into a percolating network:** *(a) Polarons can lead to real doping by forming a percolating network:* Localized polarons promote insulating behavior unless they can percolate to form a continuum. Using DFT with Eq. (4), Liu et al.[212] observed that DFT calculations of n-type doping of the Nd$_2$CuO$_4$ Cuprate superconductor at a low doping concentration of 6.25% (Fig. 20a), predicted an insulating phase due to the formation of an electron polaron with energy in the band gap, localized on the Cu site and accompanied with local lattice distortion. The wavefunction of this intermediate band is localized, and at this low concentration it does not percolate to overlap other similar states (Fig. 20b,c). When the doping concentration reaches 12.5% (Fig. 20d), the Cuprate becomes a band conductor with substantial polaron overlap (Fig. 20e,f). These findings are in qualitative agreement with the conductivity measurement of Ce-doped Nd$_2$CuO$_4$ and Pr$_2$CuO$_4$, showing a semiconductor-metal–superconductor transition at ~14% n-type doping concentration[213,214]. This illustrates how polarons (intermediate bands that are localized) can form a doping agent of a host compound and creates an insulator-to-metal transition as a function of concentration.

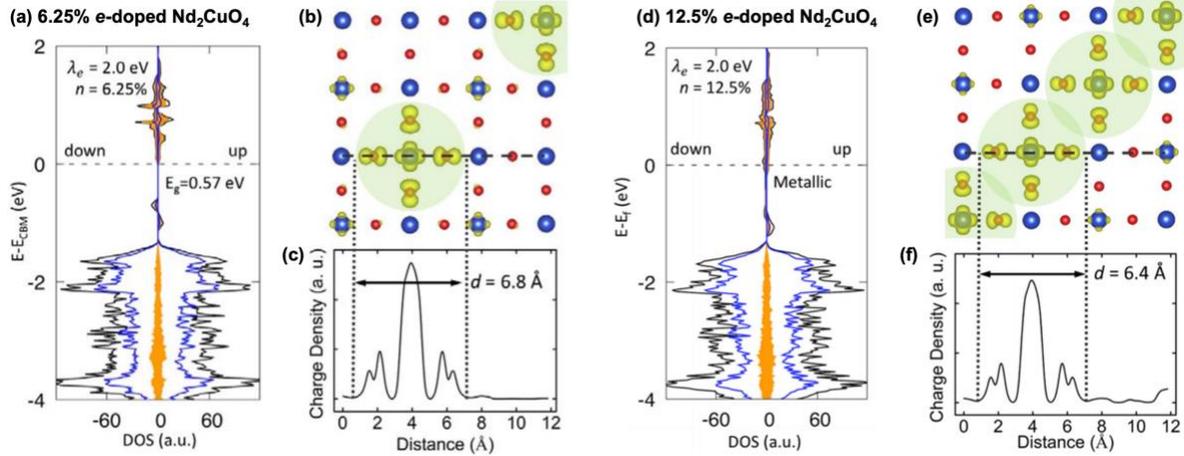

**Figure 20.** Density of states (DOS) (black) and projected DOS of O and Cu (orange and blue, respectively) of $Nd_2CuO_4$ for (a)-(c) 6.25% and (d)-(f) 12.5% electron-doping. Charge density for the highest occupied states below $E_F$ for(b) 6.25 % and (c) 12.5% electron-doping showing polaronic localization (green circles). At low doping concentration, the system is an insulator with DFT band gap energy of 0.57 eV, while for the higher doping concertation the polaronic states overlap, resulting in a metallic system. (c,f) One-dimensional charge density along the dotted line with corresponding polaron diameters for different doping concentrations. The figure is reprinted with permission from Ref. [212], Copyright (2018) by the American Physical Society.

***(b) Can creation of vacancies having local magnetic moments lead to collective ferromagnetism in common nonmagnetic insulators?*** The scenario of polaron percolation forming a conductive state can also occur for localized impurity states. This subject arose because of the prospect of magnetism due to doped vacancies *in nonmagnetic oxides*, i.e., magnetism without magnetic ions.[102-108] Consider, for example, a *charge-neutral* cation vacancy in divalent monoxides such as MgO and CaO: it supports a two-hole defect center, whereas the *singly negative vacancy* has one hole in the gap level, and a *doubly negative vacancy* has a fully occupied level without holes. Likewise, the cation vacancy in four-valent oxides such as $HfO_2$ creates a four-hole center for the charge-neutral vacancy and a fully occupied level for the quadruple negative $V_{cation}$. The open shell charge states may give rise to local magnetic moments. The question is whether the vacancy-vacancy interactions of centers with local moment can lead to collective ferromagnetism as perceived in Ref [102-108]. This suggestions has been examined by Osorio et al for CaO[110] and for $HfO_2$.[109] In CaO, the isolated charge-neutral vacancy $V^0(Ca)$ has a local magnetic moment due to its open-shell structure. The ferromagnetic vacancy-vacancy pair interaction can be calculated by DFT by contrasting the total energy of structures having two vacancies separated by distance R with the total energy of two isolated vacancies. It turns out that in CaO this V-V interaction extends only to four neighbors or less.[110] To achieve magnetic percolation on a fcc lattice with such an interaction range one needs a minimum of a concentration $1.8 \times 10^{21}$ $cm^{-3}$. Total-energy calculations for CaO[110] show, however, that due to the high vacancy formation energy even under the most favorable growth conditions one cannot obtain more than 0.003% or $10^{18}$ $cm^{-3}$ vacancies at equilibrium, showing that a nonequilibrium vacancy enhancement factor of X 1000 is needed to achieve collective magnetism in such systems. Similarly, it was previously noted theoretically[104] that charge-neutral Hf vacancies in $HfO_2$ have partially occupied electronic levels, and thus the vacancies can carry a non-vanishing local magnetic moment. DFT total energy calculations [109] indicated that pairs of such vacancies interact ferromagnetically if they are separated by up to third-neighbor distance. One might then inquire if such vacancies can lead to collective ferromagnetism. By calculating the energy required to form such vacancies in $HfO_2$ as a function of the chemical potential and Fermi level, it was possible to compute, as a function of growth temperature and oxygen pressure, the equilibrium concentration of those vacancies that have a local magnetic moment. This shows that under the most favorable equilibrium growth conditions the concentration of Hf vacancies does not exceed $6.4 \times 10^{15}$ $cm^{-3}$ (a fractional composition of $2 \times 10^{-7}$ %). Ref.[109] gives the (Monte Carlo–

calculated) percolation thresholds of various lattices symmetries as a function of the percolation radius of the interaction. Using such universal percolation graphs, it turns out that the minimum Hf vacancy concentration needed to achieve wall-to-wall percolation in the $HfO_2$ lattice, given the range of the magnetic V(Hf)-V(Hf) interaction extending to five neighbors is $x_{eq}$=13.5% under the most favorable growth conditions. Thus, itinerant ferromagnetism can be established only if one beats the equilibrium Hf vacancy concentration during growth by as much as eight orders of magnitude.

**5.10 Electronic defects: Antidoping of polaron-like states:** In systems that already have a polaron or a split-off intermediate band one might wonder how would such electronically induced states behave if deliberately doped by added carriers. While traditional electron (hole) doping shifts the Fermi level to the conduction (valence) band, some quantum materials can also exhibit a very unusual type of doping ("antidoping"). This occurs in bulk compounds that already have a split-off intermediate band that traps carriers; such split-off bands are often mistaken for proper valence band (if occupied; see Fig. 18 for $YTiO_3$ and $LaTiO_3$ as an example) or conduction band (if empty; see Figs. 3 and 21 for $YNiO_3$ as an example). Doping such systems n-type (p-type) will results in shifting the previously *unoccupied* intermediate band (*occupied* intermediate bands) to the principal valence (conduction) band[111-118] just in the reverse direction of traditional doping. In the concentrated limit, such doping results in band gap opening and increase of resistivity. Liu et al.[116] have recently demonstrated that *electron* antidoping is a general effect that can be observed in the family of compounds having (i) unoccupied intermediate bands that (ii) contain trapped holes, i.e., usually localized on the ligands. For instance, $YNiO_3$ is predicted to be a compound that can show antidoping behavior.[118] This compound has a hole trapped intermediate band, which is localized on both Ni and O atoms (Fig. 21a,b). Upon doping of $YNiO_3$ by 1e per Ni, the previously structurally and electronically inequivalent Ni atoms (Fig. 21b) become identical to each other (Fig. 21c), and band gap opening is observed.

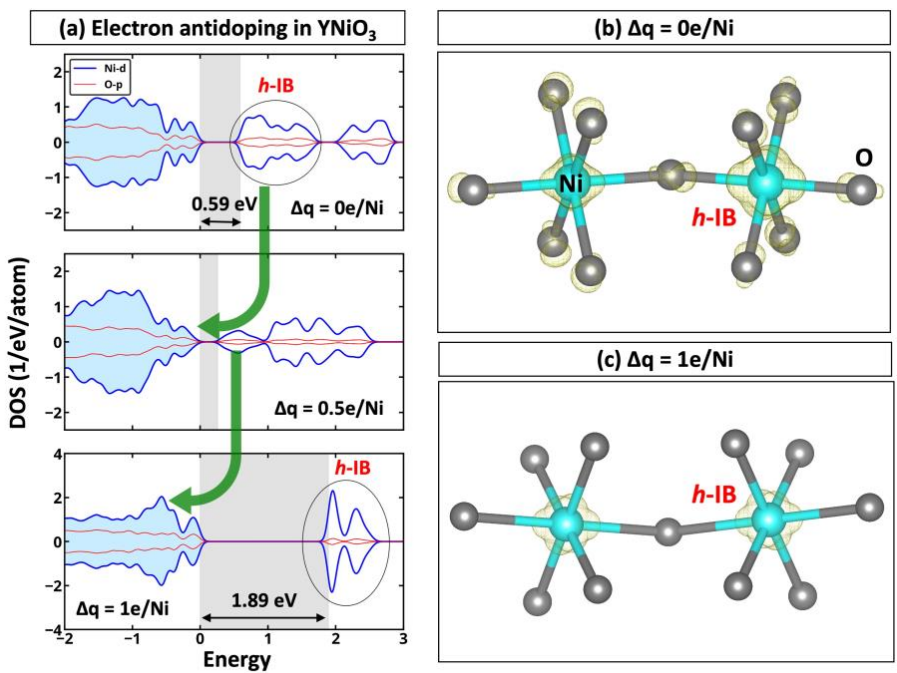

**Figure 21.** Electron antidoping in $YNiO_3$. (a) Density of states for undoped and electron-doped $YNiO_3$ for different doping concentrations. Wavefunction squared of the hole-trapped intermediate band (h-IB) shown in yellow for (b) undoped and (c) 1e/Ni-doped $YNiO_3$. Doping the system by 0.5e/f.u. shifts a part of the *h*-IB towards the principal valence band, as illustrated by the green arrow. Doping $YNiO_3$ by 1e/f.u results in the band gap opening. The band gap of the system is shown in silver. Occupied states are shadowed in light blue. The figure is reprinted with permission from Ref.[118], Copyright (2020) by the American Physical Society.

*Hole* antidoping is predicted instead in compounds having electron trapped intermediate bands usually localized on reduced cations. This behavior is demonstrated on the example of $Ba_2Ti_6O_{13}$ (see Fig. 22), which is an insulator with in-gap states occupied by 2e/f.u. Hole doping of this compound results in oxidation of the reduced cation (e.g., transformation of $Ti^{3+}$ to $Ti^{4+}$), resulting in band gap opening at certain doping concentration, e.g., when all reduced cations are converted. In general, the hole antidoping is expected for early transition or rare earth metal oxides having composition weighted formal oxidation state larger than zero.[118] While antidoping is a relatively new discovery, from the fundamental perspective, it complements the textbook understanding of doping chemistry, which can be used to design new multifunctional materials that can accumulate large concentration of free carriers and whose properties can be governed via controllable doping.

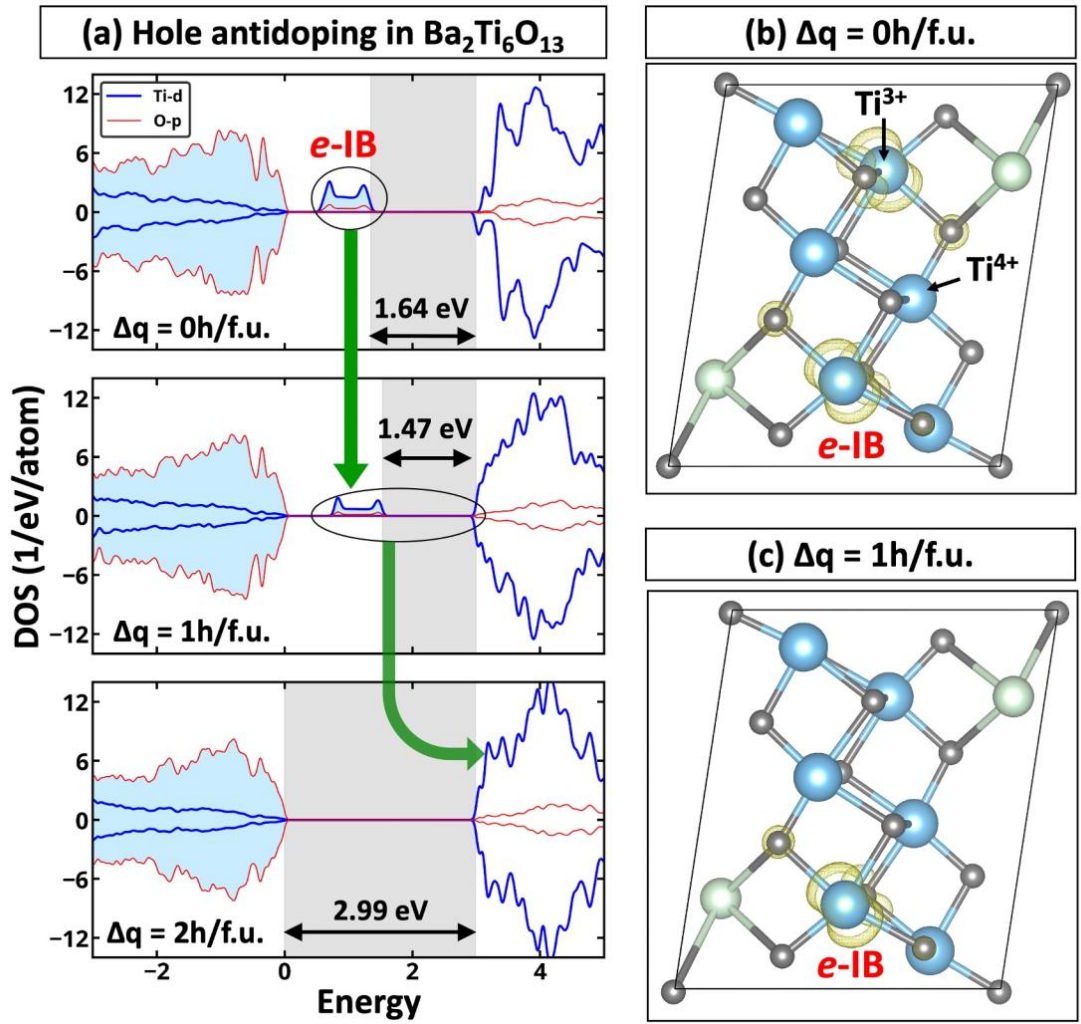

**Figure 22**. Hole antidoping of $Ba_2Ti_6O_{13}$. (a) Density of states of pure and hole-doped $Ba_2Ti_6O_{13}$. Adding the hole to the system converts a single reduced "$Ti^{3+}$" to "$Ti^{4+}$", reducing the population of the electron-trapped intermediate band (e-IB) by moving the doped bands to the principal conduction band as illustrated by the green arrow. The further hole doping results in band gap opening. The wavefunction squared of the e-IB is shown in yellow for (b) the undoped and (c) 1h/f.u. doped $Ba_2Ti_6O_{13}$. Ti, O, and Ba atoms are marked by light blue, gray, and light cyan, respectively. The band gap of the system is shown in silver. Occupied states are shadowed in light blue. The figure is reprinted with permission from Ref. [118], Copyright (2020) by the American Physical Society.

**5.11 The presence of polarity in films can change the thermodynamics of defects and doping:** One of the exciting developments in quantum materials has been the observation of 2DEG at interfaces of bulk insulators: At a critical thickness of LaAlO$_3$ grown on top of SrTiO$_3$, interfacial 2DEG is observed, despite the fact that the two components of the heterojunction are insulators in bulk.[60] LaAlO$_3$ is a polar structure that consists of alternating (LaO)$^+$–(AlO$_2$)$^-$ layers. Whereas bulk LaAlO$_3$ has a large formation energy of oxygen vacancies[215], making the concentration of oxygen vacancies V(O) in bulk very low, the polarity of LaAlO$_3$ *film* reduces the formation enthalpy (Eq. (1)) of oxygen vacancies, so that above a critical film thickness this energy vanishes.[61] Application of defect theory to surfaces showed that the formation energy ∆H of the surface vacancies decreases linearly as the film thickness of LaAlO$_3$ increases.[216,217] The surface oxygen vacancy has its donor levels located energetically above the SrTiO$_3$ conduction band minimum; the ionized vacancy electrons are then swept by the intrinsic field to the interface. Thus, it is the polarity that triggers thermodynamically the spontaneous formation of certain defects that, in turn, cancel the polar field induced by the polar discontinuity. The ionization of the spontaneously formed surface oxygen vacancy, rather than the defect-free band bending, is responsible for the conductivity and magnetism at the interface. This dependence of defect formation energy on polarity, as well as the existence of any other surface donors with transition energy above the CBM of the underlying heterojunction partner (SrTiO$_3$), offers yet another knob for controlling (interfacial) doping of quantum materials.

**5.12 Interstitial hydrogen doping leads to a universal doping pinning level:** Doping of organic solids is traditionally done by introducing interstitial alkali metals[8-10] that shed their valence electrons if the alkali outer s level lies above the host CBM, creating free electrons. In the inorganic world, this has been done by hydrogen insertion[11,15,16]. In most cases, hydrogen forms an interstitial impurity in such systems and can exist in three charge states H$^+$, H$^0$, and H$^-$. Considering the H formation energy vs. parametric Fermi level (Fig. 23a), one sees that the charge transition level $\varepsilon(+/-)$ involve a two-electron jump from H$^+$ to H$^-$, known as an Anderson Negative U system[218]. The interesting observation[12] is that this $\varepsilon(+/-)$ level occurs in some host compounds above their CBM, thus instilling electron conductivity, as in SnO$_2$ and ZnO[11,16], whereas in other compounds $\varepsilon(+/-)$ occurs in the band gap as a deep level, leaving the system insulating. Kilic and Zunger generalized[12] such observations by calculating $\varepsilon(+/-)$ for a few compounds by DFT, noting that if the VBM and CBM band edges of different compounds are approximately aligned with each other using the accepted band offset values (Figs. 15 and 16), then $\varepsilon(+/-)$ tends to stay constant in different materials when refereed to this common reference point (Fig. 23b). This "universal" doping pinning level E$_{pin}$ was generalized by van de Walle and Neugebauer[13] to include more compounds (Fig. 23c). This principle, analogous to the pinning concept predicted in Fig. 15 and 16 separated hydrogen dopable compounds into two groups: the one where interstitial hydrogen impurity is expected[12] to forms a donor such as SnO$_2$, CdO, ZnO, Ag$_2$O, HgO, CuO, PbO, PtO, IrO$_2$, RuO$_2$, PbO$_2$, TiO$_2$, WO$_3$, Bi$_2$O$_3$, Cr$_2$O$_3$, Fe$_2$O$_3$, Sb$_2$O$_3$, Nb$_2$O$_5$, Ta$_2$O$_5$, FeTiO$_3$, and PbTiO$_3$, whose conduction band minimum lie below this E$_{pinn}$, i.e., will become conductive once hydrogen is incorporated into the lattice. Conversely, materials such as MgO, BaO, NiO, SrO, HfO$_2$, and Al$_2$O$_3$, whose CBM lie above this pinning level, will remain non-conductive. Van der Walle and Neugebauer[13] extended this result to III-V semiconductors (Fig. 23c), finding that hydrogen does not dope Si, SiC, AlN, GaN, GaP, InP, GaAs but would dope InN n-type and, GaSb and InSb both p-type. The realization of the existence of the interstitial dopant pinning level unifies knowledge in organic and inorganic systems and might find application in novel quantum materials when certain doping type is sought.

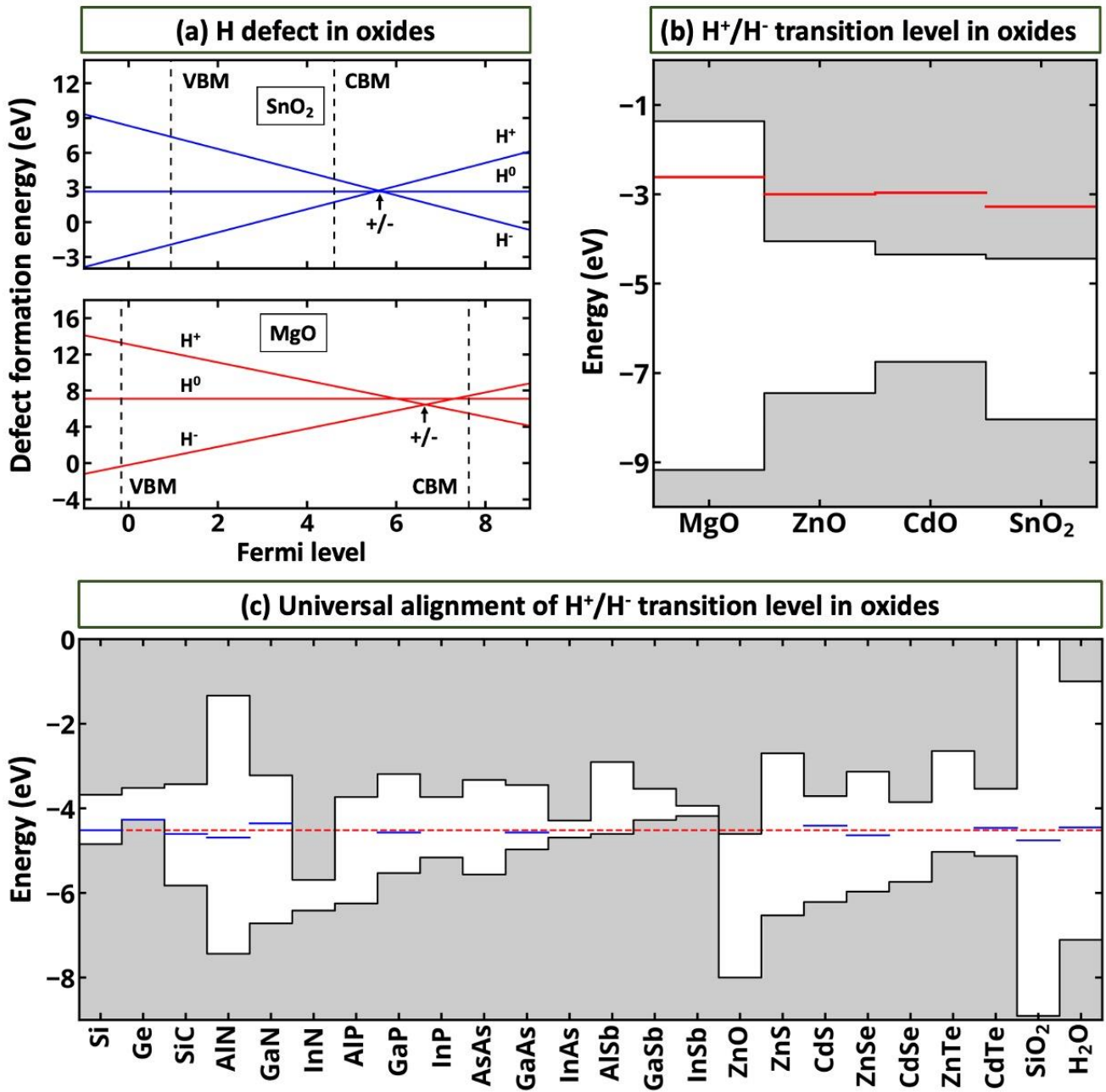

**Figure 23.** (a) Formation energy of interstitial hydrogen in $SnO_2$ and MgO as a function of the Fermi level. (b) $H^+/H^-$ transition level for different oxides presented with respect to their band offsets. The figure is redrawn from Ref. [12]. (c) Band offsets and alignment of $H^+/H^-$ transition level for different oxides. The figure is redrawn from Ref. [13].

## 6. SUMMARY AND CONCLUSIONS

The "modern theory of doping" has been developed over the past years in the sub-field of semiconductor physics, where a long history of focused experimental studies propelled by the rapidly growing electronic and photonic technologies has provided detailed experimental data to be theoretically explained. The emergence of "Quantum Materials", where uniquely quantum interactions between the components produce specific behaviors, is now pointing attention to a range of doping-related phenomena associated with different chemical classes, including ternary and multinary wide-gap oxides containing open-shell d electrons, or heavy element compounds. The electronic structure of *undoped* 3d oxides has been traditionally described by "highly correlated" methods based on the central role of on-site interelectronic repulsion in a symmetry-preserving picture. This view was motivated in part by the perceived failure of mean-field like (i.e., featuring a single effective potential in the underlying Schrodinger equation) density functional theory. Such failures motivated the introduction of dynamic electron-electron correlation in a symmetry-preserving picture, as an essential, enabling ingredient in treating such compounds. A more recent examination (summarized in Sec. 3) showed that the DFT that was said to fail was based on the smallest number of possible magnetic, orbital, and structural degrees of freedom, i.e., a rather naïve (N) version of DFT. It turns out that the N-DFT methodology is not all that actual DFT can do, and there are avenues for removing the constraints on N-DFT other than disposing of DFT altogether. Examinations of what are the minimal, enabling physics concepts needed to understand the trend in the properties of undoped 3d oxides across the Periodic Table suggests that mean-field-like DFT, with energy lowering spin- and space- symmetry breaking, (Sec. 3) is sufficient. This approach requires that one use DFT with large unit cells and symmetry unrestricted wavefunctions so that different symmetry breaking modalities able to lower the total energy are allowed. This leads to systematic understanding in terms of structural, orbital, and magnetic symmetry breaking, all sanctioned by the DFT, mean-field framework, without appeal to dynamic correlation. This includes specifically effects previously described in the highly correlated literature as exclusively enabled by dynamically correlated physics, absent in the view of this literature from DFT (actually, from naïve density functional theory (N-DFT)). The exceptions to this success pertain to physical phenomena where explicit multiplets survive covalency, as is the case for in-gap d-d transitions in wide-gap d-electron compounds, or heavy, f-electron metals showing the Kondo overlap of multiplets with band states. The current work reviewed here on doping 3d oxide and other quantum materials is therefore largely based on symmetry, spin, and space unrestricted DFT, using large supercells, allowing various forms of atomic displacements or spin organizations.

Many of the phenomena encountered in doping quantum materials fly in the face of a few naïve expectations based on experience with doping Si or Ge. It turns out that the quantum materials are not "electronically rigid", when doped, but instead are capable of rearranging structurally in Le Chattelier-like response to the introduction of carriers. In other words, doping creates antibodies for doping, thereby limiting the range over which the Fermi level can be tuned by doping (Sec. 5.1). In the old days, it was believed that wide-gap insulators cannot be doped at all. It now appears that many of those that can be doped but can do so only one way—either by electrons (n-type) or by holes (p-type), demonstrating a fundamental asymmetry (Sec. 5.2). Indeed, both the self-regulating (homeostasis like) response, creating structural defects with opposing charges to those intended in deliberate doping, as well as doping induced phase transitions can limit the range where Fermi energies can be tuned with impurity (Sec. 5.3). Interestingly, the understanding of such trends led to the development of Pauling-esque rules that systematize such as "doping rules", suggesting strategies for "designer doping" (Sec. 5.4). Deviations from the ideal expectation of perfect Daltonian stoichiometry are not only systematic but also lead to "natural doping" (Sec. 5.5), destroying important quantum properties (Sec. 5.6) or offering control of the generally contraindicated functionalities of transparency and conductivity, much needed in optoelectronics (Sec. 5.7). An interesting development in this regard is that in addition to the traditional *atomic doping,* (i.e., by impurities creating a perturbation potential $\Delta V_{imp}=V(I)-V(H)$), one can recognize *electronic impurities* whereby say, the $Ti^{4+}(d^0)$ would electronically reconstruct into $Ti^{3+}(d^1)$, creating an electronic perturbation $\Delta V_{elec}= V(Ti^{3+}) - V(Ti^{4+})$, capable of splitting a sub-band from the broad parent conduction band into the band gap region (Sec. 5.8). The creation of deep gap isolated bands trapping carriers (polaron) can dope some Cuprates (Sec. 5.9) or lead to the interesting effect of "antidoping" (Sec. 5.10) whereby doping by electrons (holes) a compound manifesting such in- gap states can shift the Fermi level towards the principal valence band (conduction band), as seen in lithiation of $Li_xFeSiO_4$, $Li_xIrO_3$, and in n-type doping of $SmNiO_3$ and $YNiO_3$ rather than in the conventional reverse direction. Finally, we note two interesting cases where doping and defect formation can be drastically enhanced by the *polarity of the compound,* as in spontaneous oxygen vacancy formation in $LaAlO_3$ associated with the interfacial conductivity of SrTiO$_3$/LaTiO$_3$ interfaces (Sec. 5.11), and the emergence of a universal demarcation point $\varepsilon(+/-)$ separating conductive from the insulating effect of doped interstitial hydrogen (Sec. 5.12).

There are numerous challenges in understanding doping in quantum materials associated with the fact that such compounds are generally not electronically rigid, affording instead significant reorganization of both atomic and electronic structure in response to doping. The good news is that electronic structure modalities based on DFT allow the atomic positions, spin configurations, unit cell symmetries to change self consistently, creating a *feedback loop* between local structure and electronic properties. In contrast, fixed Hamiltonian band structure methodologies such as standard tight binding or other model Hamiltonians lack a feedback mechanism that allows the electronic structure and the crystal or spin structure to affect each other. Instead, the fixed local atomic and spin structure end up deciding the outcome of electronic characteristics uniquely. It is this highly nonlinear feedback characteristic of responsive electronic structure methods (such as DFT) that is at the heart of understanding the peculiarities of quantum materials and their doping, more so than possible electron-electron correlation corrections to a mean-field representation of interelectronic interactions

## AUTHOR INFORMATION

**Corresponding author**

Alex Zunger – Energy Institute, University of Colorado, Boulder, Colorado 80309, United States; Email: alex.zunger@colorado.edu.

**Authors**

Oleksandr I. Malyi – Energy Institute, University of Colorado, Boulder, Colorado 80309, United States.

**Notes**

The authors declare no competing financial interest.

**Biographies**

*Prof. Alex Zunger* is a condensed matter theorist of real materials at the Energy Institute, University of Colorado, Boulder [group web page http://www.colorado.edu/zunger-matter-by-design]. Ph.D. Tel Aviv University, Israel (Joshua Jortner and Benjamin Englman); Postdoc Northwestern Physics (A.J. Freeman) and Berkley Physics (Marvin Cohen). Recipient of the Rahman Award of the American Physical Society, the (inaugural) "Materials Theory Award" of the Materials Research Society, the John Bardeen award of the TMS, the "Tomassoni Prize" (Italy), the Medal of the Schola Physica Romana, and the Boer Medal for Photovoltaic research. Involved in foundational developments of theoretical methods (exchange-correlation functional, first-principles pseudopotentials, momentum space total energy and plane-wave band theory) in Density Functional Theory, quantum nanostructures, theory of alloys, defects and surfaces in a variety of solid classes, and more recently in materials by design with focus on the Inverse Design approach.

**Oleksandr I. Malyi** is a solid-state physicist at the Energy Institute, University of Colorado Boulder. He received his Ph.D. in materials science from Nanyang Technological University in 2013. After postdoc training in Singapore and Norway, he joined Prof. Alex Zunger's group as research associate. His main research expertise is in theoretical inverse design of solid-state materials, doping physics, and understanding of materials in general.

## ACKNOWLEDGMENT

A.Z. acknowledges with enormous gratitude the collaborations over the years with his post and Research Associates in the research area of doping and defects: Sergey Barabash, Marilia Caldas, Adalberto Fazzio, Hiroshi Katayama-Yoshida, Cetin Kilic, Qihang Liu, Stefan Lany, Ulf Lindefelt, Priya Mahadevan, Jorge Osorio-Guillén, Tula Paudel, Clas Persson, Hannes Raebiger, Vladan Stevanovic, Liping Yu, Zhi Wang, Su-Huai Wei, and Seng Bai Zhang. The work at the University of Colorado at Boulder was supported by the U.S. Department of Energy, Office of Science, Basic Energy Sciences, Materials Sciences and Engineering Division, under Grant No. DE-SC0010467 to the University of Colorado. The *ab initio* calculations of this work used resources of the National Energy Research Scientific Computing Center, which is supported by the Office of Science of the U.S. Department of Energy.